\documentclass[twocolumn,aps,prb]{revtex4} 
\usepackage{graphics,amsmath,graphicx} 

\begin{document}
\title{Electronic transport and quantum localization effects
in organic semiconductors.}

\author{S. Ciuchi} 

\affiliation{Dipartimento di Scienze Fisiche e Chimiche\\ 
Universit\`a dell'Aquila, CNISM and Istituto Sistemi Complessi CNR, 
via Vetoio, I-67010 Coppito-L'Aquila, Italy} 

\author{S. Fratini} 

\affiliation{Institut N\'eel - CNRS \& Universit\'e Joseph Fourier \\ 
BP 166, F-38042 Grenoble Cedex 9, France}

\begin{abstract}

We explore the charge transport mechanism in organic semiconductors based  on a
model that accounts for the 
thermal intermolecular disorder at work in pure crystalline compounds,  
as well as extrinsic sources of disorder that are 
present in current experimental devices.
Starting from the Kubo formula,
we describe a theoretical framework that 
relates the time-dependent quantum dynamics of electrons to the
frequency-dependent conductivity.
The electron mobility is then calculated through a relaxation time
approximation that  accounts for quantum localization corrections
beyond Boltzmann theory, and  
allows us to efficiently address 
the interplay between highly conducting states in the band range and 
localized states induced by disorder in the band tails.
The emergence of a ``transient localization'' phenomenon
is shown to be a general feature of organic semiconductors, 
which is compatible with the 
 bandlike temperature dependence
of the mobility observed in pure compounds.
Carrier trapping by extrinsic disorder causes a crossover to a
thermally activated behavior at low temperature, which is
progressively suppressed upon increasing the carrier concentration,
as is commonly observed in organic field-effect transistors. 
Our results establish a direct connection between the localization of the
electronic states and their conductive properties,  
formalizing phenomenological considerations that are commonly used in the
literature.

\end{abstract}

\date{\today}
\maketitle
\section{Introduction}

Remarkable progress has been made in recent years 
in  understanding and improving 
electronic transport in organic semiconductors (OSC)
and devices. 
Mobilities exceeding $10cm^2/Vs$ are now  measured in
an increasing number of organic field-effect transistors (OFETs) based
on single crystals \cite{Podzorov,Xie,Minder,Sakanoue,Liu}.
Such values are orders of magnitude lower than those attainable
in inorganic semiconductors, and are indicative of extremely short
electronic mean-free paths ---
on the order of 
the inter-molecular distances \cite{Friedman,Cheng} --- causing a 
breakdown of the basic assumptions underlying band transport.
This occurs despite the relatively modest coupling of the carriers with
intra-molecular vibrations, which rules out the presence of polarons in such
materials.\cite{orgarpes11,Stojanovic}
It is currently believed  
that the mobility in crystalline organic semiconductors 
is intrinsically limited by  the presence of
large thermal molecular motions,
which are a direct consequence of the weak Van
der Waals inter-molecular bonds. 
\cite{Munn,Troisi06,Picon,reconcile09,RTA11}
Deviations from the perfect crystalline arrangement
 act as a dynamical source of  disorder on the already narrow electronic
bands arising from the $\pi$-intermolecular overlaps, 
inducing
a  localization of the electronic wavefunctions on the timescale
of the inter-molecular vibrations themselves\cite{RTA11} ---
a phenomenon that is not described by
 the  semiclassical Boltzmann theory of electron-phonon scattering, nor 
 by the classical Marcus electron transfer theory. 
Theories based on such ``transient electron
localization''  \cite{Troisi06,Picon,RTA11}
are able to explain the power-law decrease of the mobility with
temperature observed 
in ultrapure organic
semiconductors \cite{Karl} as well as the optical
conductivity data available in OFETs \cite{Li,Fischer,RTA11}.

Experimentally, the intrinsic mobility of organic semiconductors
is still difficult to observe in practical OFET
devices. Even when polaronic self-trapping \cite{Hulea} and dipolar 
disorder\cite{Richards08} induced by the interface polarizability 
are avoided by using non-polar gate dielectrics or suspended samples, 
the carrier mobility in OFETs is still affected 
by extrinsic sources of disorder,
related  to the presence of structural defects or 
to the interface roughness. 
Extrinsic disorder favors the formation of trapped states in the band
tails,  
at  energies located below those of the intrinsic carriers \cite{Kalb,Xie}. 
As a result, depending on the device quality,
a crossover from an intrinsic transport regime  
to a thermally activated (trapped) regime is
observed upon lowering the
temperature \cite{Xie,Podzorov,Minder}, or the intrinsic
regime can be completely washed out if the disorder is sufficiently strong as 
occurs in polycrystalline films. \cite{Chang11}

As is clear from the above discussion, 
a proper description of the transport mechanism in both
pure and disordered OSC requires 
a method that (i) goes beyond both Boltzmann and Marcus approaches and (ii) 
is able to describe  the interplay between highly
conducting  states in the band range 
and weakly mobile states induced by disorder.
This is achieved here by applying a recently developed
theory of charge transport 
based on the Kubo formula for the electrical conductivity
 combined with a suitable  
relaxation time approximation (RTA) on the current-current correlation
function, which takes quantum localization effects into account.
\cite{Mayou00,RTA11}
The present method 
  has already been successfully applied to analyze the quantum transport
properties of quasicrystals\cite{Mayou00,Trambly06} and to address
 the role of defects in
graphene\cite{Trambly11}. 
It has also been shown to provide an efficient
description of  the transient localization phenomenon in
pure OSC\cite{RTA11}, as it gives access to the {\em time-resolved}
diffusivity and localization length of electronic states. 
By addressing the  same quantities
{\em resolved in energy}, we show here that
this theoretical framework also
establishes a direct relationship between the existence of
competing electronic states at different energy scales and the
resulting transport properties. 
Accordingly, both the intrinsic transport mechanism of clean organic
semiconductors and the crossover to 
a thermally activated motion in the presence of extrinsic disorder are
rationalized in terms of the relative weight played by
strongly localized tail states 
and more mobile electronic states in the band range.
The increase of the mobility observed in OFETs
upon injecting a sufficiently large density of carriers is also
naturally explained within this scenario.

The paper is organized as follows.  In Sec. \ref{sec:general} we introduce the 
formalism relating the quantum diffusion
of electrons
to the Kubo response theory.  Based on this formalism,
in Sec. \ref{sec:approx} we briefly describe the semiclassical approximation
used in Ref. \onlinecite{reconcile09} and then 
derive the relaxation time approximation to be used here.
A model relevant to organic semiconductors and devices 
is introduced in Sec. \ref{sec:model}.
The results obtained  in the
limit of low carrier concentration are presented in  Section
\ref{sec:results} and their density-dependence   is analyzed in
Sec. \ref{sec:density}. The main conclusions are drawn in Sec. \ref{sec:conclusions}.

\section{General formalism}
\label{sec:general}

A formalism that relates the quantum diffusion
of electrons, i.e. the quantum mechanical spread of the electron position with 
time,  to the optical conductivity was introduced in
 Refs. \onlinecite{Mayou00,Trambly06} for metals and generalized to semiconductors  in Ref. 
 \onlinecite{RTA11}. 
The main steps of the derivation are reviewed
here. Readers not interested in formal developments may skip
this  Section and move on 
directly to Section III.

\subsection{Optical conductivity and time-resolved diffusivity}
We start from the  Kubo formula that relates the response of
electrons to an
oscillating electric field to the current-current correlation
function (say, along $x$)\cite{Mahan}:
\begin{equation}
  \label{eq:Kubostandard}
  \sigma(\omega)=\frac{1}{\Omega  \omega} Re \int_0^\infty dt
  e^{i(\omega + i\delta)t} \langle [J_x(t),J_x(0)]\rangle.
\end{equation}
Here $\delta$ is a small positive number enforcing convergence,
$\Omega$ is the system volume, and we have set $\hbar=1$.
Denoting the retarded current-current correlation function as
\begin{equation}
  \label{eq:Cpm}
  C_-(t)= \theta(t)\langle [\hat J_x(t),\hat J_x(0)]\rangle ,
\end{equation}
and its its Fourier transform as $ C_-(\omega)$,
the Kubo formula Eq. (\ref{eq:Kubostandard}) can be expressed as
\begin{equation}
  \label{eq:Kubocompact}
  \sigma(\omega)=\frac{1}{\Omega \omega} Re \ C_-(\omega).
\end{equation}

A relation between the mean square particle displacement and the current
correlations can now be obtained through the retarded current-current  
anti-commutator correlation function,
\begin{equation}
C_+(t)= \theta(t)\langle \lbrace\hat J_x(t),\hat J_x(0)\rbrace \rangle.
\label{eq:retardedCp}
\end{equation}
Writing the current operator  
in terms of the velocity operator, $\hat J=e\hat V=e d\hat X/dt$,  
and performing the time derivative
we see that this function is directly related to the mean square
displacement $\Delta X^2(t) = \langle |\hat X(t)-\hat X(0) |^2
\rangle$ of the total position operator $\hat X(t)=\sum_i \hat x_i(t)$
along the chosen direction, 
(with $\hat x_i$ the  position operator for the $i$-th particle) via 
\begin{equation}
  \label{eq:vv}
  \frac{d \Delta X^2(t)}{dt}=\frac{1}{e^2}\int_0^t C_+(t^\prime) dt^\prime.
\end{equation}
This defines the instantaneous diffusivity of a system of $N$
quantum particles,
\begin{equation}
  \label{eq:diffusivity}
  {\cal D}(t)=\frac{1}{2}\frac{d \Delta X^2(t)}{dt}.
\end{equation}


Introducing the 
mean square displacement  reached by the $N$-particle system over
a typical timescale $\tau$  as
\begin{equation}
\label{eq:defL2}
L^2(\tau) = \int_0^\infty dt e^{-t/\tau}  \frac{d \Delta X^2(t)}{d t},
\end{equation}
and using the properties of  Laplace trasforms of derivatives, 
Eq. (\ref{eq:vv}) yields the following relation between the 
mean square displacement
and the Laplace transform  $C_+(p)$ of the
anti-commutator correlation function,  
\cite{symbols}
\begin{equation}
\label{eq:Laplace}
 C_+(p=1/\tau)=e^2\frac{L^2(\tau)}{\tau}.
\end{equation}
The above equation shows that the quantity $C_+(p)$  has a precise physical 
meaning: it corresponds to  the diffusivity of
the electronic system averaged on a timescale $\tau=p^{-1}$.


Because the functions $C_+$ and $C_-$ are
related by the detailed balance condition, which in Fourier space reads
\begin{equation}
\label{eq:detbal}
Re C_-(\omega)=\tanh \left(\frac {\beta \omega}{2}\right) Re C_+(\omega)
\end{equation}
(with $\beta$ the inverse temperature, 
see   appendix \ref{app:detbal}), the two relations  
Eqs. (\ref{eq:Kubocompact}) and (\ref{eq:Laplace}) are 
are not independent.
Indeed, by expressing
the right-hand side of Eq. (\ref{eq:Kubocompact}) in terms of the Laplace
 transform $C_+(p)$, via Eq. (\ref{eq:CpCw}), 
we obtain an expression relating  the
mean square displacement  $L^2(\tau)$
and the optical conductivity $\sigma(\omega)$:
\begin{equation}
  \label{eq:L2omega}
  L^2(\tau)=\int_0^{\infty} \frac{d\omega}{\pi}
  \frac{2\omega}{\omega^2+(1/\tau)^2}\frac{\sigma(\omega)}{\tanh(\beta
    \omega/2)}.  
\end{equation}
Remarkably, this relationship allows to
address the time-resolved diffusion of electrons from the knowledge 
of the optical conductivity, which is  a spectral property. 
An analogous equation was derived in Ref. \onlinecite{RTA11} for the
instantaneous spread $\Delta X^2(t)$. 


\subsection{d.c. conductivity and mobility}
From the equivalence of the two formulations
Eqs. (\ref{eq:Kubocompact}) and (\ref{eq:Laplace})
we can 
derive a generalized Einstein relation connecting the electrical
conductivity to the extensive  diffusion coefficient ${\cal D}$, 
which is valid for quantum
$N$-particle systems.
By definition, a system is diffusive if
  the diffusivity at long times tends to a
constant value, $\lim_{t\to \infty} {\cal D}(t)={\cal D}$.
In the limit
$\tau \to \infty $  
the integral in Eq. (\ref{eq:defL2}) is then dominated
by such asymptotic diffusive behavior leading to 
\begin{equation}\label{eq:asympt}
{\cal D}=
 \lim_{\tau \to \infty } \frac{L^2(\tau)}{2\tau}= \frac{C_+(p=0)}{2e^2}.
\end{equation}
Conversely, the above equation  shows that reaching a 
finite localization length $L(\tau\to \infty)$  in the long time limit 
implies a vanishing diffusion coefficient.

Using Eq. (\ref{eq:Laplace}) together with the definition
of the d.c. conductivity from Eq. (\ref{eq:Kubocompact}) as the limit
\begin{equation}
  \label{eq:sigmadc}
  \sigma=\lim_{\omega\to 0} \frac{Re C_-(\omega)}{\Omega \omega}
\end{equation}
and observing that $\lim_{p\to 0}
C_+(p)=\lim_{\omega\to 0} C_+(\omega)/2$ (see appendix \ref{app:detbal})   
we can write
\begin{equation}
  \label{eq:Einstein}
  \sigma=\frac{e^2}{k_BT\Omega}{\cal D }=\frac{e^2}{2k_BT\Omega}{\lim_{\tau \to \infty } \frac{L^2(\tau)}{\tau}}.
\end{equation}

Our definition of the extensive 
diffusion coefficient  ${\cal D}$ for the N-particle system  differs from
the usual single particle diffusivity, which we denote
$D$. The latter 
is an intensive quantity,  defined as the
ratio between the conductivity and the 
charge-charge
susceptibility \cite{Kubo57},  so that 
\begin{equation}
  \label{eq:relationKK}
  \frac{\cal D}{\Omega}=D  \frac{\partial n}{\partial \mu}  k_BT
\end{equation}
with $n$ the density and $\mu$ the chemical potential. 
The density-dependent proportionality factor on
the r.h.s.
 corresponds to the number of particles that can actually
move, i.e. the compressibility times the (thermal) energy
interval. 
Accordingly, the mobility of electrons   can be defined at any finite
density via 
\begin{equation}
  \label{eq:mobfin}
  \mu_e=\frac{eD}{k_B T}=\frac{\sigma}{e k_B T  \frac{\partial n}{\partial \mu}}. 
\end{equation}


\subsection{Energy resolved quantities}
 \label{sec:energy-resolved}

Our aim is to address the charge dynamics in systems
where localized and itinerant states coexist  in 
 different regions of the electronic spectrum: tail states generated by
disorder below the band edges  behave differently
from states within the electronic band. It is therefore useful
to
decompose the response of the electronic
system into contributions from states at different energy
scales.\cite{reconcile09,Coropceanu12}  
This can be done by exploiting the following 
expression of $C_-(\omega)$ as an
 energy integral (see appendix \ref{app:detbal}):
\begin{equation} 
  \label{eq:Kuboexpand}
   C_-(\omega)=\pi\int d\nu \left[ f(\nu)-f(\omega+\nu)\right] tr
   [\hat \rho(\nu) \hat J \hat \rho(\nu+\omega) \hat J],
\end{equation}
where and $f(\nu)=[e^{\beta(\nu-\mu)}+1]^{-1}$ is
the Fermi function,
\begin{equation}
  \label{eq:defrho}
   \hat \rho(\nu)=-\frac{1}{\pi} Im \frac{1}{\nu-\hat H}
\end{equation}
is the spectral operator from which  the DOS $\rho(\nu)$ can be
obtained as $\rho(\nu)=tr\hat \rho(\nu)$, 
and $\hat H$ is the Hamiltonian operator. 
Defining 
\begin{equation} 
  \label{eq:corrJJ}
  B(\nu)= tr [\hat \rho(\nu) \hat J \hat \rho(\nu) \hat J],
\end{equation}
the d.c. conductivity is readily obtained from Eq. (\ref{eq:sigmadc})
as 
\begin{equation}
  \label{eq:Kubo-Fourier}
  \sigma=\frac{\pi}{ \Omega} \int d\nu B(\nu) 
\left(-\frac{\partial f}{\partial\nu} \right)
\end{equation}
[note that there was a misprint in the definition of $B(\nu)$ in
Ref. \onlinecite{reconcile09}].
We see from the above equation 
that the 
total conductivity  of an electronic system arises from an average of $B(\nu)$
over all electronic
states, weighted by the corresponding statistical population. 
For example,
in a system at finite electron density and low temperature,
because the derivative of the Fermi function is peaked at
$\nu \simeq \mu$, the conductivity is determined by the electrons in proximity
(within $k_BT$) of  the chemical
potential,  leading to $\sigma\simeq (\pi/\Omega)B(\mu)$. 

We now show that 
$B(\nu)$ is actually proportional to the energy-resolved
diffusivity of states at energy $\nu$.
In the case of independent electrons which is of interest here,
the electron mobility can be evaluated at any finite density 
via Eq. (\ref{eq:mobfin}), using the following expression for    
the compressibility
\begin{equation}
  \label{eq:compr}
  \frac{\partial n}{\partial \mu}
=\int d\nu \left(-\frac{\partial f}{\partial\nu} \right)\rho(\nu).
\end{equation}
From Eqs. (\ref{eq:mobfin}) and (\ref{eq:Kubo-Fourier}), 
and defining  the 
diffusivity of states at energy $\nu$ as 
\begin{equation}
\label{eq:resolvD}
 D(\nu)=(\pi/e)[B(\nu)/\rho(\nu)].
\end{equation}
we can rewrite the mobility  as 
\begin{equation}
  \label{eq:mobdens}
  \mu_e=\frac{e}{k_BT}\frac{\int d\nu \left(-\frac{\partial f}{\partial\nu} \right)
   \rho(\nu) \ D(\nu)}{\int d\nu \left(-\frac{\partial f}{\partial\nu} \right) \rho(\nu)},
\end{equation}
which has explicitly the form of an average over energy with 
a probability distribution $W(\nu)= \left(-\frac{\partial f}{\partial\nu}
\right) \rho(\nu)/\int d\nu \left(-\frac{\partial f}{\partial\nu} \right) \rho(\nu)$.
In the  limit of vanishing density, by taking the
 $\mu\to -\infty$ limit appropriate to a non-degenerate
 semiconductor in Eq. (\ref{eq:mobdens}), we find 
\begin{equation}
  \label{eq:mobzero}
  \mu_e=\frac{e}{k_BT}\frac{\int d\nu
    \rho(\nu)e^{-\beta \nu}D(\nu)}{\int d\nu \rho(\nu)e^{-\beta \nu}} .
\end{equation}
%



\section{Approximation schemes}
\label{sec:approx}






\subsection{Semiclassical Kubo bubble  approximation}

A powerful approximation scheme to calculate the carrier
mobility is to evaluate the Kubo formula using
the exact electron propagators obtained in the limit of static molecular
displacements, but neglecting  
vertex corrections \cite{reconcile09}.
Evaluating the single-particle propagators in the static limit is 
justified in virtue of the  low frequencies of the
intermolecular vibrations that couple to the electronic motion.
On the other hand, 
the neglect of vertex corrections amounts to dropping  the quantum
interference processes that are responsible for Anderson
localization,\cite{LeeRMP,vertex} 
in the spirit of the semiclassical
approximation. 
It corresponds to replacing
 the function $B(\nu)$ appearing in Eq. (\ref{eq:mobdens})  by the 
factorized expression
\begin{equation}
B(\nu)= tr \left \langle \ \langle\hat \rho(\nu)\rangle \ \hat J \
  \langle\hat \rho(\nu) \rangle \ \hat J \ \right\rangle,
\label{eq:BubbleNoVtx}
\end{equation}
where $\langle \ldots \rangle$ 
means an average over  disorder variables
(the averaging
procedure will be
defined in the following Section).
In diagrammatic terms, only the elementary particle-hole  ``bubble'' ---  a
convolution of two spectral functions ---  is retained in the
evaluation of the current-current correlation function. 

  
While Eq. (\ref{eq:BubbleNoVtx}) neglects 
particle-hole quantum correlations,
it still accounts for
non-trivial interaction effects contained in the 
single-particle propagators, which are calculated exactly.
In particular, it is able to capture
those aspects of the transport mechanism 
which stem  from the dual nature of the  
 electron states.
It is therefore superior to the 
usual Bloch-Boltzmann treatment in that it can account for both the
coherent motion of  band
states and the incoherent motion of tail states. 
The convolution Eq. (\ref{eq:BubbleNoVtx}) is actually
analogous to the form that applies in the limit of large lattice
connectivity underlying dynamical mean field theory,
and that has proven successful to address
the crossover from band-like to hopping motion of small polarons
non-perturbatively.\cite{Fratini03}  
It reduces to the  static treatment of the gaussian
disorder model presented in Refs. \onlinecite{Bassler} and 
\onlinecite{Coehoorn} in the classical limit where the
electron bandwidth is neglected, which is
appropriate for narrow-band amorphous semiconductors and polymers 
in the strong disorder regime.


\subsection{Relaxation time approximation (RTA)}

The theoretical framework developed in Sec. II allows us to restore the 
backscattering processes leading to Anderson localization,   
i.e. those that are neglected in Eq. (\ref{eq:BubbleNoVtx}), 
by performing a
physically transparent relaxation time approximation (RTA)
\cite{RTA11,Mayou00,Trambly06}.
The idea underlying the RTA is to express the dynamical properties of
the electronic system under study in terms of those of a suitably defined
reference system  from which it decays over time, 
and that can be solved at reasonable cost. 
In the semiclassical theory of electron transport, for example,
one starts from a perfectly periodic crystal and describes 
via the RTA the decay of momentum states due to the scattering by impurities or
phonons.
The idea here is to find an alternative
reference system to start with, so that quantum 
localization effects are built-in from the beginning.
We now show that this can be achieved by 
starting from the exact description of a 
``parent'' localized system where the  disorder is assumed to be
static. The RTA can then be used to
restore the disorder dynamics related to  
the low-frequency lattice
vibrations.\cite{RTA11,Mayou00,Trambly06}

Let us consider an organic semiconductor where 
the disorder variables (i.e. the molecular positions) 
fluctuate in time over a typical timescale $\tau_{in}$.
At times  $t\ll \tau_{in}$,
the  molecular lattice appears to the moving electrons as an
essentially frozen disordered landscape. 
The velocity correlation function $C_+(t)$ [cf. Eq. (\ref{eq:Cpm})]
then coincides with what would be 
 obtained if the
disorder were static, 
which we denote $C^{loc}_+(t)$ (our reference system). 
In particular, the buildup of quantum interferences
underlying Anderson localization --- which 
occurs on the scale of the {\it elastic} scattering time $\tau_{el}$ ---
is realized provided that  $\tau_{el}<\tau_{in}$. In this time range,
the organic semiconductor therefore exhibits all the features of a truly
localized electronic system.
Quantum interferences that were present in the parent localized
system are instead destroyed at longer times because, due to the
lattice dynamics, the electrons encounter different disorder
landscapes when moving in the forward and backward directions \cite{LeeRMP}.
The form
\begin{equation}
   \label{eq:defRTA}
   C_+(t)= C^{loc}_+(t) e^{-t/\tau_{in}}
\end{equation}
is the simplest form that is able to capture such decay process. 
Transforming Eq. (\ref{eq:defRTA}) to Laplace space, results in
$C_+(p)=C_+^{loc}(p+1/\tau_{in})$. 

The corresponding diffusion coefficient can be straightforwardly obtained 
from  Eq. (\ref{eq:asympt}). 
We see that
starting from a localized system with a
vanishing diffusion coefficient, $C_+^{loc}(p\to 0)=0$,
the RTA  restores a finite
diffusion coefficient 
which equals the diffusivity of the localized system at a time
$\tau_{in}$.  This result can be expressed as
\begin{equation}
\label{eq:diffRTA}
{\cal D}=\frac{L^2_{loc}(\tau_{in})}{2\tau_{in}},  
\end{equation} 
 which is analogous to the
Thouless diffusivity of Anderson insulators \cite{Thouless}.  
Correspondingly, the quantity $L^2_{loc}(\tau_{in})$ 
evaluated through Eq. (\ref{eq:defL2}) for the reference localized system
acquires the meaning of a {\it transient localization length} for the
actual dynamical system, 
as it represents the
typical electron spread achieved after the initial localization stage,
and before diffusion sets back in at $t>\tau_{in}$ 
(see Figs. 1 and 3 in
Ref. \onlinecite{RTA11} as well as Fig. 1 below 
for a real-time illustration of this behavior).
The emerging physical picture of the electronic motion that follows
from the RTA Eq. (\ref{eq:diffRTA})  is quite
different from the usual semiclassical picture, where disorder and
lattice vibrations cause rare scattering events on extended
electronic states. 
The present scenario rather describes 
electrons that are prone to localization but can take advantage of the
dynamics of disorder to diffuse freely over a distance 
$L_{loc}(\tau_{in})$, with a trial rate $1/\tau_{in}$. 
As will be shown in Sec. V,
 the RTA  essentially reproduces 
the results obtained from more time-consuming 
mixed quantum-classical simulations \cite{Troisi06,RTA11,Wang}, and is free
from the known drawbacks of these approaches.

Before presenting model-specific results in Section \ref{sec:results}, 
we analyze in more detail how the energy-resolved quantities 
of Sec. \ref{sec:energy-resolved} translate into the  RTA language. 
From Eqs. (\ref{eq:mobfin}) and (\ref{eq:diffRTA}) the RTA mobility in
the low density limit is 
\begin{equation}
   \label{eq:defmuRTA}
   \mu_e = \lim_{n\rightarrow 0} \frac{e}{n
     k_BT\Omega}\frac{L^2_{loc}(\tau_{in})}{2\tau_{in}}. 
\end{equation}
In the spirit of Eq. (\ref{eq:mobzero}),  
the transient localization length $L_{loc}(\tau_{in})$ can 
be expressed in terms of its energy resolved equivalent, 
$\ell(\tau,\nu)$, i.e. 
the spread reached {\em by electronic states of energy $\nu$} 
at time $ \tau_{in}$,  
as
\begin{equation}
\label{eq:Lextensive}
\lim_{n\rightarrow 0}\frac{L^2_{loc}(\tau_{in})}{\Omega n} =
\frac{\int d\nu \rho(\nu) e^{-\beta \nu} \ell^2(\tau_{in},\nu)}{\int
  d\nu \rho(\nu) e^{-\beta \nu}}  
\end{equation}
[see Appendix \ref{sec:BoltzmannRTA} for an explicit expression of
$\ell^2(\tau_{in},\nu)$] \cite{nota-ell-Kubo}.
Combining Eqs. (\ref{eq:mobzero}), 
(\ref{eq:defmuRTA}) and (\ref{eq:Lextensive}) 
we recognize the energy-resolved diffusivity 
\begin{equation}
  \label{eq:RTAbub}
   D(\nu)=    \frac{ \ell^2(\tau_{in},\nu)}{2\tau_{in}}
\end{equation}
which  relates directly 
the conduction properties of the electronic states to 
their localization length in the parent localized system.

Finally, from the considerations of the preceding Section  we can 
derive the following relation:
\begin{equation}
  \label{eq:relmuRTAsigma}
  \mu_e = \frac{e}{2\tau_{in}k_BT}
  \int_0^{\infty} \frac{d\omega}{\pi}
  \frac{2\omega}{\omega^2+\tau_{in}^{-2}}
  \frac{\sigma_{loc}(\omega)/n}{\tanh(\beta\omega/2)}.  
\end{equation}
Eq. (\ref{eq:relmuRTAsigma}) expresses the electron mobility in the RTA 
in terms of the optical conductivity of the reference localized system, whose
mobility strictly vanishes. This result 
deserves a few comments.
From  scaling theories of localization,  a finite 
d.c. conductivity is customarily obtained by  
taking the optical conductivity  
 to saturate at a  cutoff frequency of the order of
 the inverse of the inelastic scattering time, 
$\sigma_{d.c.}= \sigma_{loc}(\omega\simeq\tau_{in}^{-1})$. 
In  Eq. (\ref{eq:relmuRTAsigma}), instead,
  the inelastic scattering time enters into the determination of
  the mobility  via a
  weighted integral (i.e. through a lorentzian convolution) 
  that involves the conductivity at all frequencies.
The mobility  Eq. (\ref{eq:relmuRTAsigma}) 
can therefore be quite different from the value obtained
from the usual thumbrule.
Our scheme
is also conceptually different from the
approach used in Ref. \cite{Cataudella}. There 
the mobility was obtained by performing a Lorentzian convolution
of the optical conductivity itself, with a phenomenological
broadening $\Gamma=1/\tau_{in}$ 
that was assumed to originate from the {\it quantum}
fluctuations of the molecular vibrations instead of the {\it classical} 
molecular motions. 
Apart from its different physical content, 
the method of Ref. \onlinecite{Cataudella} provides, 
for a given value of $\tau_{in}^{-1}$, 
a lower estimate for the mobility than Eq. (\ref{eq:relmuRTAsigma}).


\section{Model and method}
\label{sec:model}

We now apply the theoretical framework developed in the preceding Sections to 
 a model relevant to organic semiconductors,
that accounts for both the  intrinsic dynamical
disorder caused by 
inter-molecular motions \cite{Troisi06,reconcile09,RTA11,Cataudella} and the
fluctuations of the molecular site energies that are assumed to 
 originate from  
extrinsic sources disorder.
Specifically, we consider the following tight-binding Hamiltonian, for
electrons or holes on a one-dimensional molecular
lattice 
\begin{equation}
  \label{eq:H}
H=  \sum_i \epsilon_i c^+_i c_i +\sum_{\langle i j\rangle} t_{ij}
c^+_i c_j + h.c.
\end{equation}
where $\epsilon_i$ are molecular site energies, and $t_{ij}$
are intermolecular transfer integrals between nearest neighboring
molecules. In a perfect crystal all site energies $\epsilon_i$ are
equal and can be set to zero without loss of generality. Static disorder
leads to variations of the site energies with 
a  statistical distribution $P(\epsilon_i)$. 
We are interested here in the effects of energetically distributed disorder, 
as opposed to disorder centers with a definite energy (such as specific
point defects). Correspondingly, we take $P(\epsilon_i)$ 
to be a gaussian of variance $\Delta$ as a representative case study.
In addition, the coupling of the electrons to the vibrations of the molecules
 induces a dynamical disorder in
the inter-molecular transfer integrals $t_{ij}$ which depend on the
molecular positions $R_i,R_j$. These fluctuate 
on a timescale
governed by the relevant  vibrational modes, whose frequency we denote
as $\omega_0$
\cite{Troisi06,reconcile09,RTA11}.  
 We assume a linear dependence of
$t_{ij}$ on the intermolecular distance, $t_{ij}=t_0
[1-\alpha(R_i-R_j)]$.

One-particle properties --- i.e. the properties that derive from
the electron Green's function, such as the spectral function, the
quasiparticle lifetime
or the density of states (DOS) --- can be efficiently evaluated by treating the
molecular degrees of freedom as static.
This is justified  because 
the frequencies of the 
inter-molecular vibrations that couple to the electron motion  are 
much smaller than 
the band energy scale, which follows from 
the large molecular mass. For example in rubrene\cite{TroisiAdv}
$\omega_0\simeq 4-9meV$, $t_0\simeq 130meV$, so that $\omega_0\ll t_0$ 
(see \cite{Troisi06,Hannewald,WangJCP07} for different compounds). 
The static approach amounts to
treating the positions $R_i$ as classical
variables distributed according to a Gaussian
distribution of thermal origin\cite{reconcile09,RTA11}  
$P_T(R_i)\propto \exp(-M\omega_0^2 R_i^2/2k_BT)$ ($M$ is the 
 molecular mass).  
\cite{reconcile09}
In practice, a numerical solution for the electronic problem is obtained 
for each given configuration 
of $\{R_i\}$  and $\{\epsilon_i\}$  and
then averaged over the disorder configurations.
The electronic properties of the model Eq. (\ref{eq:H}) in the static limit
depend on two dimensionless coupling parameters: 
$\Delta/t_0$, that  controls
the amount of  extrinsic disorder,
 and $\lambda=\alpha^2t_0/(2M\omega_0^2)$, the electron-molecular lattice coupling parameter. From the latter, 
the variance  of the intrinsic  thermal fluctuations of
the inter-molecular   transfer intergrals is obtained as 
 $s=\sqrt{8\lambda T t_0}$.

The static limit described above 
leads to a strictly vanishing particle
diffusivity for all states.
This can be seen from Eq. (\ref{eq:RTAbub}), which
 vanishes when $\tau_{in}\to \infty$. 
The dynamical nature of
the inter-molecular vibrations 
 must therefore be accounted for in order
 to address the transport properties of electrons.
Both approximation schemes described in Sec. \ref{sec:approx}
accomplish this task by  taking 
the solution of the static model as a starting point. The technical details
 are described in what follows.

(i) To obtain the spectral function $\rho(\nu)$ needed in the
semiclassical Kubo bubble (KB) approximation we adopt 
an algorithm based on regularization of the tri-diagonal
recursion formulas for the electron propagator 
(see Ref. \cite{reconcile09} for details). 
Using this method, system sizes up to 
$N=2^{16}$ sites can be achieved. 
The spectral function is then obtained after averaging up to $6 \cdot
10^5$ different realizations of the ${R_i}$ and ${\epsilon_i}$. 
The mobility is directly obtained using Eq. (\ref{eq:BubbleNoVtx}). 
The average diffusivity $C_+(p)$ is obtained by evaluating 
the optical conductivity via Eq. (A20) applying the same factorization as
in Eq. (\ref{eq:BubbleNoVtx}), and then  
using Eqs. (\ref{eq:Laplace}) and (\ref{eq:L2omega}).

(ii) To evaluate the average diffusivity $C_+^{loc}(p)$ needed in
the RTA  we use  standard exact diagonalization techniques   
on chains of up to $N=2^9$ sites. The functions
of interest are then calculated 
via their Lehman representation as a sum over the resulting
eigenstates (see Appendix A).
Since we are  considering  electrons moving in the time-dependent
potential of the fluctuating molecular lattice,
 it is  a natural
choice to associate the scale $\tau_{in}$ in the RTA Eq. (\ref{eq:defRTA}) 
with the typical timescale of inter-molecular vibrations.
Assuming a single vibrational mode with a frequency
$\omega_0=5meV$, in the range of the relevant inter-molecular vibrations 
in rubrene, we set $\tau_{in}= 1/\omega_0=10^{-13}s$.
 This assumption has been shown in Ref. \onlinecite{RTA11} to be
 consistent with the optical absorption data available in Rubrene OFETs
 \cite{Li,Fischer}. This choice is also consistent with 
the results of dynamical Ehrenfest simulations in Ref. \onlinecite{RTA11}, 
as it correctly reproduces 
the departure from localization observed at time $\tau \simeq 1/\omega_0$.
However, a more precise estimate of $\tau_{in}$ from the Ehrenfest 
results is prevented due to the inaccuracy of this 
method in the long time limit (see below).

(iii) For comparison,
we shall also present results obtained with the method of
Refs. \cite{Troisi06,RTA11,Wang}  (termed Ehrenfest in the following). 
The dynamics of the molecular positions $R_i$
are then included explicitly by adding  a vibrational term
  $H_{vib}= \sum_i  \frac{M \omega_0^2R_i^2}{2}
  + \frac{P_i^2}{2M}$ 
to the Hamiltonian Eq. (\ref{eq:H}) 
($P_i$ are the conjugate momenta of the $R_i$).
The electron diffusion is obtained via 
Ehrenfest quantum-classical dynamical simulations:  
the $R_i$ are treated as  classical variables 
subject to forces   evaluated as averages over the electronic state
obtained from the  solution of its time-dependent Sch\"odinger equation
 \cite{Troisi06,RTA11}.
We average up to
$12800$ initial conditions  on a $1024$-site chain, 
with the initial molecular displacements 
and velocities taken from the corresponding thermal distribution. For each 
initial condition we use a different set of disorder variables ${\epsilon_i}$.

(iv) By artificially freezing the ${R_i}$ variables in the simulation
 we obtain a formulation in the time domain of the static problem
 described at point (ii), that we shall refer to as Ehrenfest-S. 
From  Eqs. (\ref{eq:defL2}) and (\ref{eq:Laplace}) 
 the calculation of the time dependent mean square displacement of
Eq. (\ref{eq:diffusivity}) can then be used to obtain the quantities 
 $L^2_{loc}(\tau)$  and  $C_+(p)$ 
from a Laplace transform.


\section{Results in the low density limit}
\label{sec:results}

\subsection{Time-resolved diffusivity and transient localization}
 \label{sec:transient}
\begin{figure}[h!]
  \begin{center}
\includegraphics[width=8cm]{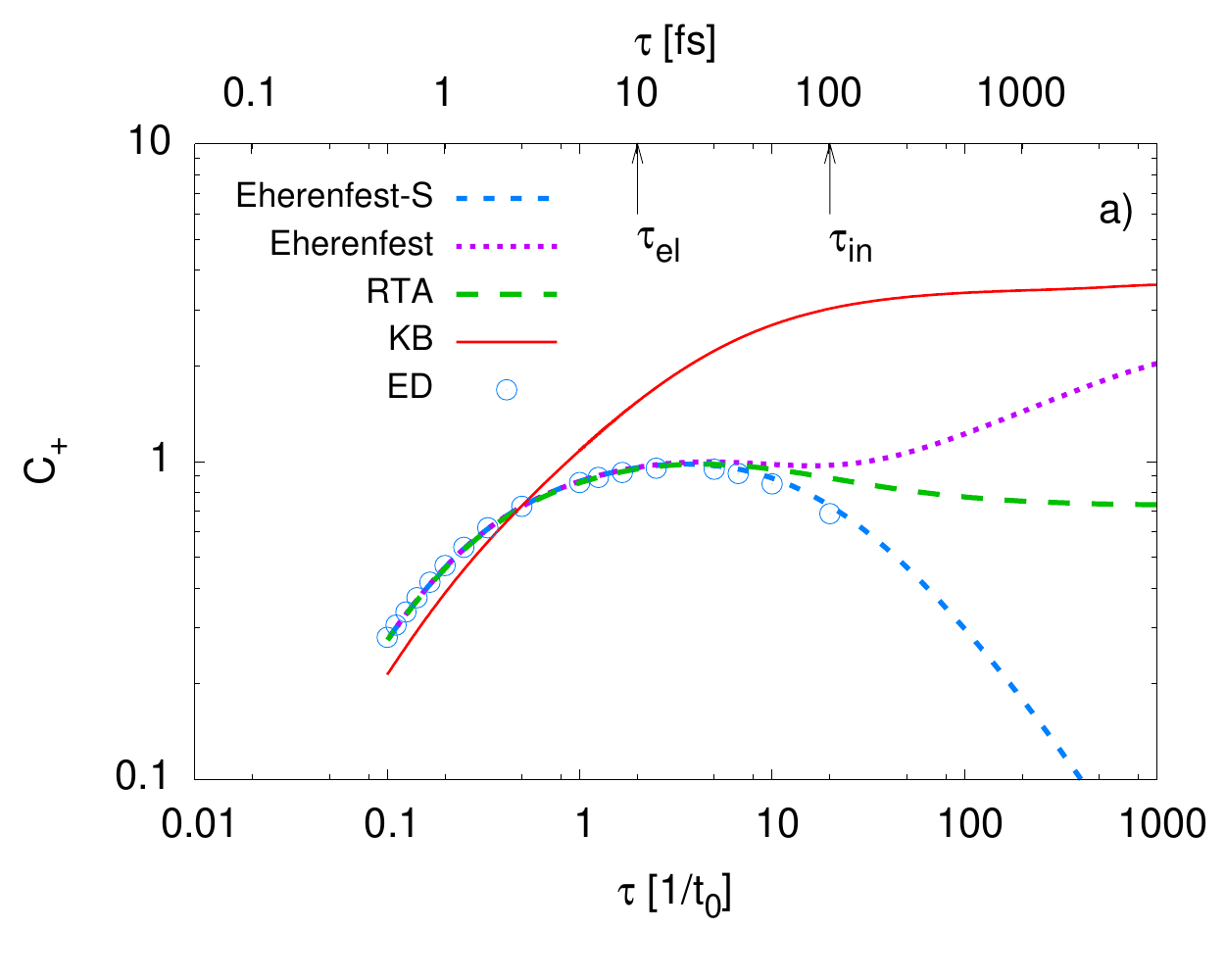}
\includegraphics[width=8cm]{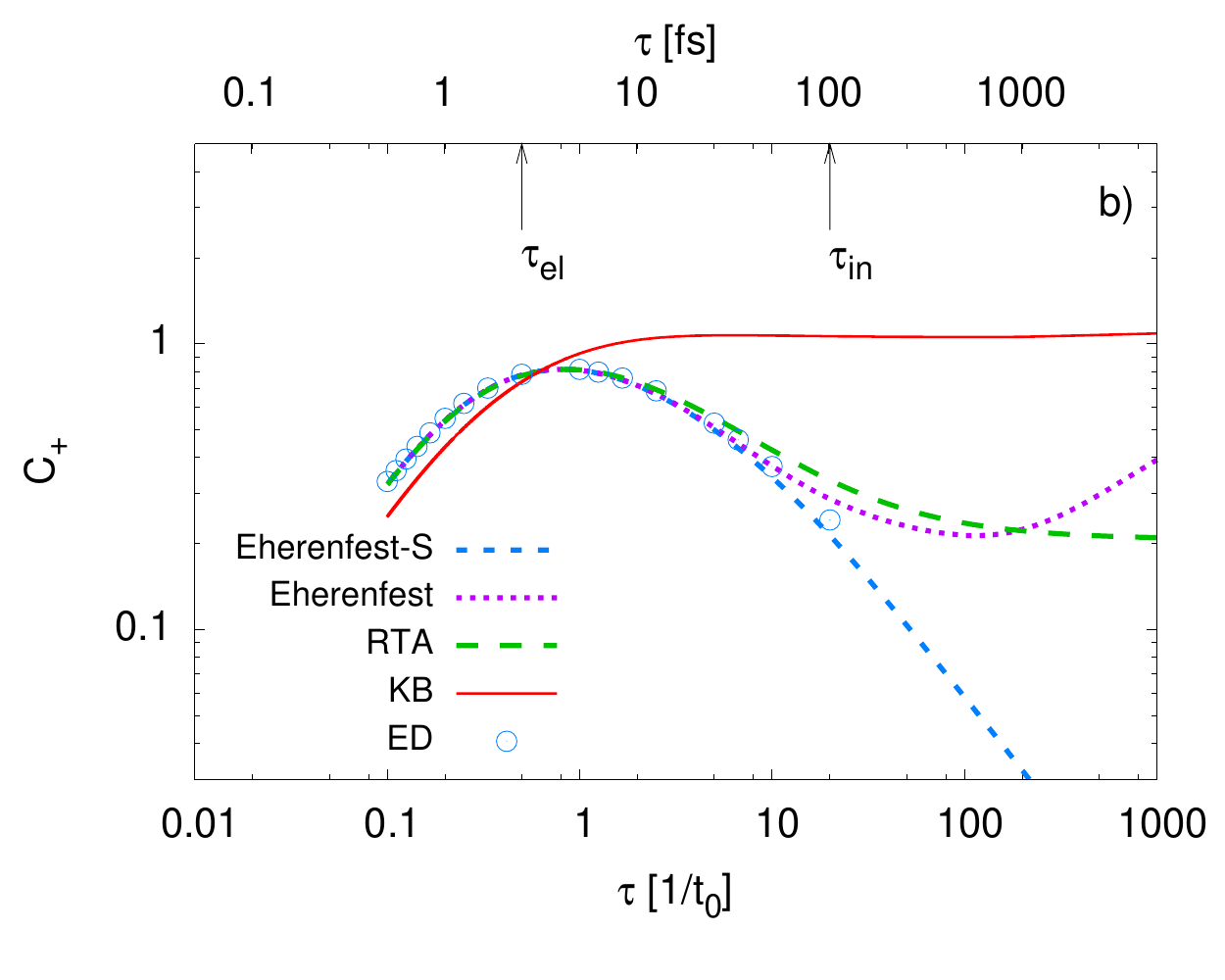}
\end{center}
  \caption{Average diffusivity $C_+(p)$ defined in
    Eq. (\ref{eq:Laplace}) as a function of the timescale $\tau=p^{-1}$, at
a temperature  $T=0.2t_0=300K$. Panels (a) and (b) are respectively 
without $\Delta=0.0$ and  with $\Delta=0.5t_0$ extrinsic disorder.
  The bold short-dashed  curve (light blue)  is obtained by solving
  the time-dependent Schr\"odinger equation on a $1024$-sites chain
  (Ehrenfest-S). 
The dotted curve (magenta)  
  is the result of the 
  dynamical Ehrenfest approach with vibrational frequency $\omega_0=0.05t_0$.
  The long-dashed line (green) is the RTA approximation 
  with $\tau_{in}=\omega^{-1}_0$.
 The full (red) line is the result of the semiclassical Kubo bubble
  approximation  Eq. 
 (\ref{eq:L2omega}).  Open circles are results from exact diagonalization
  of a $256$-sites chain.
  The arrows mark the elastic and inelastic scattering times (see text).
}
  \label{fig:CpAll}
\end{figure}

Fig. \ref{fig:CpAll} illustrates the function $C_+(p)$ defined in
Eqs. (\ref{eq:diffusivity})-(\ref{eq:Laplace}),  as calculated from
the different methods outlined at the end of the preceding Section.  
This quantity  has the
meaning of a diffusivity averaged up to  a time  $\tau=1/p$ and therefore
provides direct information on the quantum dynamics of electrons as a
function of time. 
We fix $\lambda=0.17$, which is representative for
the intrinsic electron-vibration coupling in
rubrene\cite{reconcile09}, and set the temperature to 
$T=0.2t_0=300K$. 
The numerical results for a single electron 
in a pure organic semiconductor
($\Delta=0$)  and  in the presence of extrinsic disorder
($\Delta=0.5t_0=65meV$)  are 
reported in  Figs.
\ref{fig:CpAll}a and
\ref{fig:CpAll}b respectively.  
This value of $\Delta$ is close to the one that was derived from the analysis
of angle-resolved photoemission spectra (ARPES) in crystalline pentacene 
films\cite{Hatch}.

All the methods outlined above yield a ballistic time
evolution in the short-time limit. It can be shown  that  the average
 diffusivity at
short times obeys exactly  
 $C_+= 2\langle v^2\rangle/p= 2\langle v^2\rangle \tau$, 
which is ruled by the average band
velocity $\langle
v^2\rangle$   (cf. Appendix \ref{app:sum-rules}). 
The ballistic regime is followed by a flattening
of the  average difusivity due to 
the onset of scattering processes, occurring on a timescale 
which we identify with the elastic
scattering time,  $\tau_{el}$ \cite{LeeRMP}. 
From Fig. \ref{fig:CpAll} we estimate
approximately  $\tau_{el}\sim
10^{-14} s$ in the pure case ($\Delta=0$), and $\tau_{el}\sim
5\cdot 10^{-15} s$ in the disordered case ($\Delta=0.5t_0$).

In the long time limit,
the  different methods yield qualitatively different
behaviors 
reflecting the fundamentally distinct treatments  
of the intermolecular dynamics. Let us focus on the intrinsic case first,
Fig. \ref{fig:CpAll}a.
In the parent system with  static disorder, the electrons are
localized  (Ehrenfest-S and ED, respectively 
blue short-dashed curve and open circles). The existence of
  a finite localization length $L(\tau)=const$ as $\tau\to\infty$ 
implies through Eq. (\ref{eq:Laplace}) that the average diffusivity 
bends down 
and tends to $C_+\propto 1/\tau$  at
long times. 
Restoring the dynamical nature of the
inter-molecular vibrations via the RTA 
causes a departure from the 
localized behavior on the scale of the inelastic scattering time
$\tau_{in}$,
so that a diffusive behavior  ($C_+=const$) is reached at
long times  (green long-dashed curve).
We note that within the RTA 
the diffusion coefficient at $\tau\to\infty$ is necessarily lower
than the maximum attainable value, which is obtained  when $\tau_{in}
\simeq \tau_{el}$.
 
Finally, the Ehrenfest method (purple, dotted line) also captures the
departure from localization occurring at $\tau \simeq \tau_{in}$. However,
this method 
yields a spurious superdiffusive behavior  at long times\cite{RTA11}, 
which is testified by a marked upturn of the diffusivity. 
This drawback leads to an overestimate of the
mobility, whose value can vary strongly depending on the chosen 
simulation time.  For this reason, the mobilities obtained from  Ehrenfest
simulations \cite{Troisi06,TroisiAdv,Wang,RTA11,Ishii} should 
be taken with some care.

The results reported in Fig. \ref{fig:CpAll}a, that have been obtained 
using microscopic parameters appropriate for pure rubrene (the  organic
semiconductor with the highest mobility reported to date), 
indicate that 
even in  ideal samples without extrinsic disorder the elastic
scattering time at room temperature is shorter than 
 the typical inelastic scattering time, $\tau_{el}<\tau_{in}$. 
This situation
 results from the combination of the large mass of the molecular units, 
 which leads to low vibrational frequencies and therefore to large values
 of $\tau_{in}$, together with the typically weak inter-molecular transfer
 rates and their strong sensitivity to  inter-molecular
 motions, which lead to short values of $\tau_{el}$. It is
 therefore expected to 
 be a general feature of all organic semiconductors, resulting in
 a  transport mechanism that is fundamentally different from that of
 inorganic materials.
Specifically, a  "transient" localization regime
emerges at intermediate times, $\tau_{el}<\tau< \tau_{in}$, 
where the electrons tend to localize (the
diffusivity bends down) 
 before  a constant diffusivity sets back in at
$\tau\gtrsim \tau_{in}$ (green long-dashed curve in Fig.
\ref{fig:CpAll}a).  
When such transient localization is realized, the diffusion coefficient at long
times  depends on the history of the system at this intermediate
stage, being inversely proportional to 
the inelastic time $\tau_{in}$, cf. Eq. (\ref{eq:diffRTA}). 

The existence of a transient localization phenomenon  invalidates in principle
semiclassical treatments of electron transport, which are instead
successful in inorganic materials. To illustrate this point 
we show in Fig. \ref{fig:CpAll}a the average diffusivity $C^+$
obtained from the  semiclassical Kubo bubble
approximation (red full line). Because in
this approximation the backscattering processes at the origin of
localization are neglected, the system evolves continuously from
a ballistic to a diffusive behavior. Semiclassical approaches are
therefore inadequate to describe electron transport 
in organic semiconductors where $\tau_{el} < 
\tau_{in}$.
On the other hand, 
 the present RTA is able to recover 
the semiclassical picture in the opposite regime where  $\tau_{in} <
\tau_{el}$. As discussed at the beginning of this
Section  we have that
$C^+_{loc}(p)= 2 \langle v^2 \rangle/p$ at short times. Using the RTA
Eq. (\ref{eq:defRTA}) and Eq. (\ref{eq:asympt}) yields a diffusion coefficient
$D= \langle v^2 \rangle \tau_{in}$, which is formally analogous to the
Bloch-Boltzmann result.

The situation in the presence of extrinsic static disorder 
is not qualitatively modified with respect to the pure
case,
as can be seen from Fig. 1b. In particular,  the inelastic
scattering time remains unchanged, because it is determined by the 
intrinsic timescale of the intermolecular vibrations. 
However, an increased amount of disorder shifts the
onset of localization $\tau_{el}$ to shorter times. This 
enlarges the time interval where the transient localization
phenomenon is effective, with a consequent reduction of the 
diffusion coefficient.

Finally, one can wonder if the transient localization
scenario, which was demonstrated here in the one-dimensional case, 
is robust in  more realistic descriptions of OSC.
For  materials
with a sizable in-plane anisotropy, i.e. such that the inter-molecular
transfer rates in the transverse direction are much smaller than in the
longitudinal direction, $t_\perp\ll t_0$, the one-dimensional
picture is expected to remain valid up to times $t\simeq \hbar/t_\perp$. 
This is the situation that applies to Rubrene, according to recent
photoemission data.\cite{Machida}
Because the elastic scattering time is very short, of the order of
$\hbar/t_0$ or less (see Fig. 1), we have 
$ \hbar/t_\perp \gg \tau_{el}$  and
nothing prevents the transient localization to occur in this case. 
Backscattering processes remain relevant also in isotropic
two-dimensional materials. Although the timescale for two-dimensional
weak localization is known to be longer than $\tau_{el}$, \cite{LeeRMP}
the present scenario should remain qualitatively correct also in that
case due to the
strong intrinsic disorder present in OSC.




\subsection{Temperature dependence of the  mobility}

\begin{figure}
  \centering
\includegraphics[width=8cm]{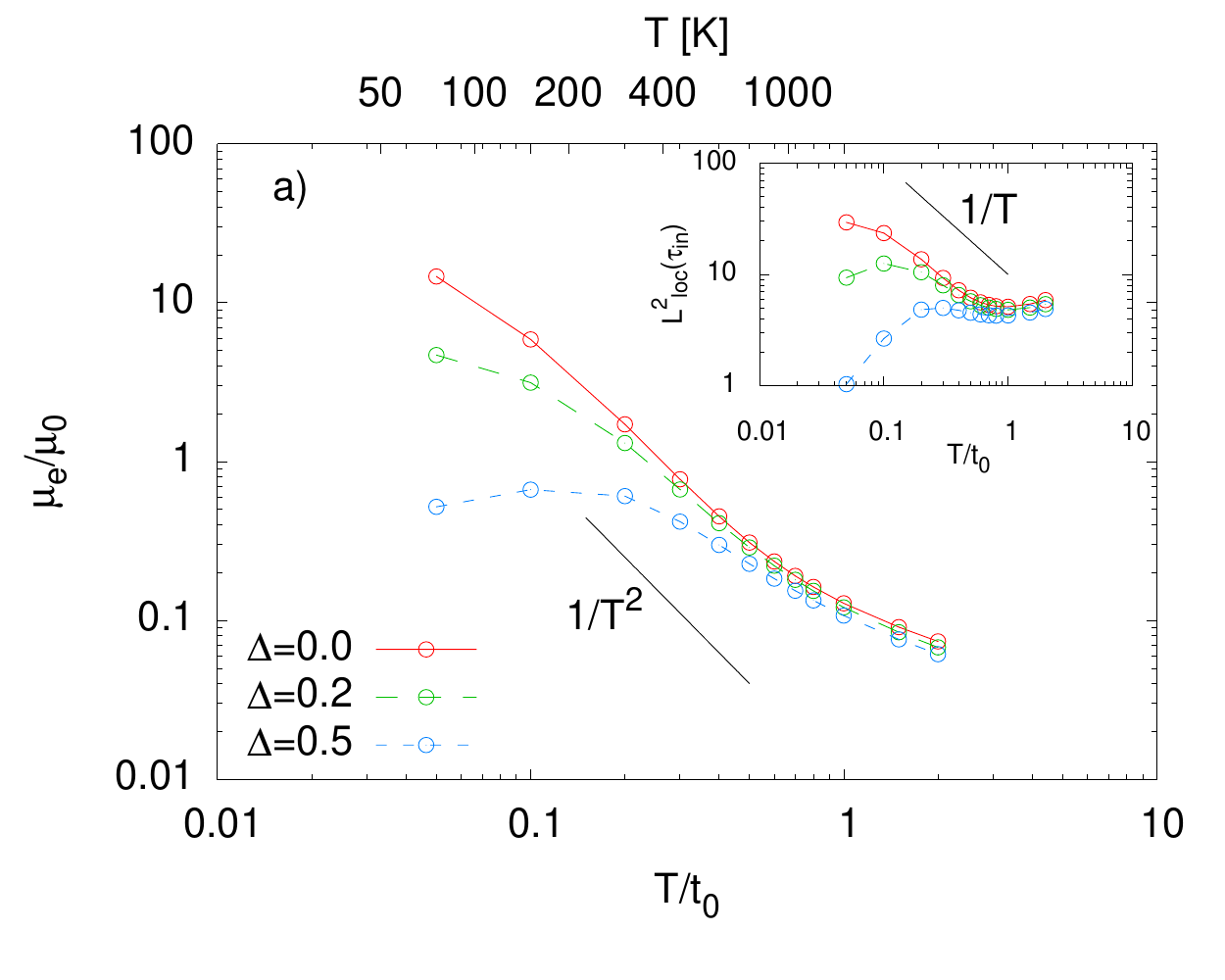}
\includegraphics[width=8cm]{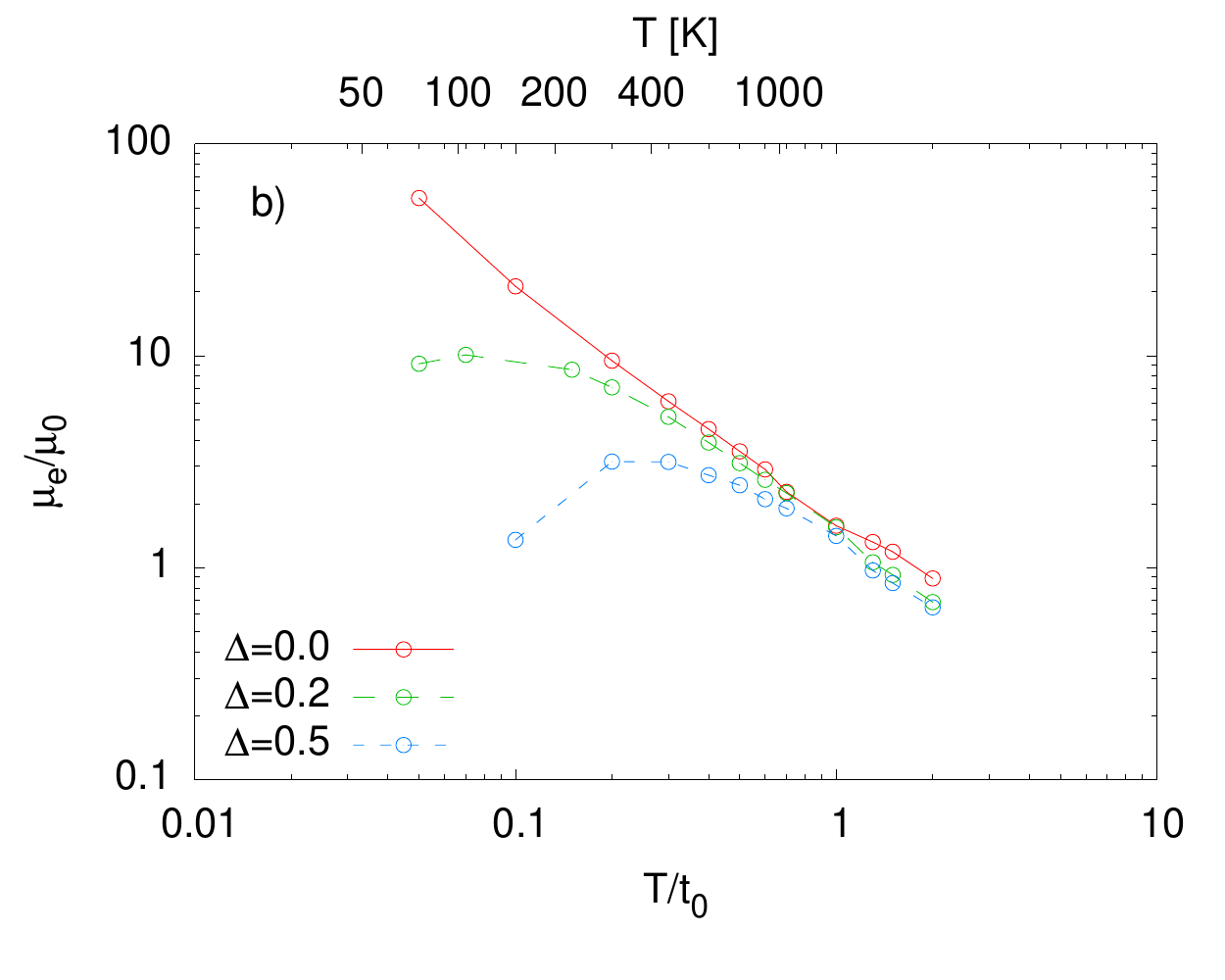}
  \caption{Mobility as a function of temperature calculated (a)  within
    the RTA 
  and (b) within the Kubo bubble approximation. The inset of panel (a)
   shows the behavior of the transient
  localization length $L^2_{loc}(\tau_{in})$ with $1/\tau_{in}=\omega_0=0.05 t_0$.
  Mobilities are expressed in units of $\mu_0=ea^2/\hbar$, with 
  $a$ the lattice spacing ($\mu_0=7cm^2/Vs$ taking $a=7.2\AA$ for 
  rubrene). In this and all subsequent figures, we take $t_0$ as the unit of energy.}
  \label{fig:mobility}
\end{figure}

Fig. 2a shows the temperature dependence of the mobility as
obtained from the RTA  
for different amounts of extrinsic
disorder, $\Delta=0$, $0.2t_0$ and $0.5t_0$.
The lowest accessible temperature is set by the limits of validity of 
our classical treatment for the molecular
vibrations, namely  $T\gtrsim 
\omega_0=0.05t_0$.
The intrinsic mobility of pure compounds 
($\Delta=0$, full red line) 
is a monotonically  decreasing function of temperature,
 even though the microscopic transport mechanism 
is far from a conventional band transport, as 
discussed in Sec. \ref{sec:transient}. 
Depending on  the explored temperature window, 
a power law temperature dependence, $T^{-\gamma}$, can be identified. 
In practice the exponent $\gamma$ depends on how the transient
localization length in Eq. (\ref{eq:defmuRTA}) varies with
temperature, due to the thermal increase of 
inter-molecular disorder.

The behavior of $L^2_{loc}(\tau_{in})$ is shown in the inset of
Fig. \ref{fig:mobility}a. 
In the temperature interval 
$T=200-600 K$, 
the transient localization length 
decreases steadily\cite{reconcile09} as $L^2_{loc}(\tau_{in})\propto
a^2 t_0/(\lambda T)$. 
Substituting this expression in Eq. (\ref{eq:defmuRTA}) leads to 
a mobility $\mu_e  \propto T^{-2}$ \ \cite{reconcile09,Troisi06}. Moreover, using explicitly the
definition of $\lambda$ given in Sec. IV and the relation $1/\tau_{in}\sim \omega_0$ 
we obtain that 
the mobility in this regime increases with the third power of $\omega_0$, 
while it is independent of the transfer integral $t_0$. An analogous calculation in
the semiclassical regime \cite{reconcile09} yields for the model under study a mobility proportional 
to  $\omega_0$ and to  $t_0^{1/2}$. In both cases, increasing $\omega_0$ results
in an increase of the charge mobility, because it suppresses the effects of dynamical 
inter-molecular disorder. This observation suggests a possible strategy for the design 
of high mobility materials: rather than optimizing the inter-molecular $\pi$ overlaps that control 
the value of $t_0$, it could be advantageous to stiffen the inter-molecular vibrations 
either via an appropriate 
tayloring of the inter-molecular structure (e.g. by molecular functionalization) 
or via the interaction with a substrate 
(as in self-assembled monolayers).

The localization length becomes a weakly  increasing function of T at very high
temperatures, where a 
vibrationally assisted electron motion arises via  the 
large fluctuations of the inter-molecular distances. This results in
a weaker power law dependence of the mobility with exponent $\gamma
<1$, which was termed ``mobility saturation'' in Ref. \cite{reconcile09}. 
Although such high temperatures are not attainable experimentally in
rubrene, the
mobility saturation regime
could actually be observed in materials with a stronger
electron-vibration coupling constant $\lambda$ than that considered here.

The inclusion of extrinsic disorder ($\Delta\neq 0$)  causes a
downturn of the mobility at low
temperatures, that is reminiscent of a thermally activated behavior,
i.e. $\mu_e$ increases with $T$. 
This behavior reflects a crossover between the extremely short
$L^2_{loc}(\tau_{in})$ obtained at low
temperatures and the larger  intrinsic value at higher temperatures,
as shown in the inset of Fig. \ref{fig:mobility}a.
\cite{note:transient}
The location of the crossover from extrinsic to intrinsic transport
depends on the   amount of extrinsic
disorder, so that  it can vary experimentally depending on the material
and device  quality.
Correspondingly, a variety of  behaviors ranging from thermal
activation to a power law decrease can be realized in the
experimental temperature window, which 
 is possibly at the origin of the different temperature
dependent mobilities observed 
in organic FETs.


The results of the semiclassical Kubo bubble approximation  are shown in
Fig. \ref{fig:mobility}b for comparison.  
Despite the profound differences in the  
two descriptions of charge transport, the overall behavior obtained in the
explored temperature interval is qualitatively  similar.
From a fundamental viewpoint, the fact that a  thermally activated  
behavior  is
obtained  for $ \Delta \neq 0$ 
in both the RTA and the Kubo bubble approach
indicates that the 
corresponding hopping processes (incoherent 
jumps from molecule to molecule) are already present at
the semiclassical level, being captured by the Kubo bubble
approximation in the strong disorder limit.\cite{Coehoorn}
This result means that the hopping behavior is not 
related to the quantum (backscattering) localization corrections, as
the latter are not contained in the semiclassical treatment.
As we proceed to show, the crossover from the intrinsic to the
thermally activated regime can be explained
in terms of the competition between highly 
conducting states located in the band range and weakly mobile states
located in the band tails ---  a competition that is 
captured by both the RTA and the Kubo bubble approximation.

\subsection{Energy resolved diffusivity and localization length}

\begin{figure*}
\centering
\includegraphics[width=8cm]{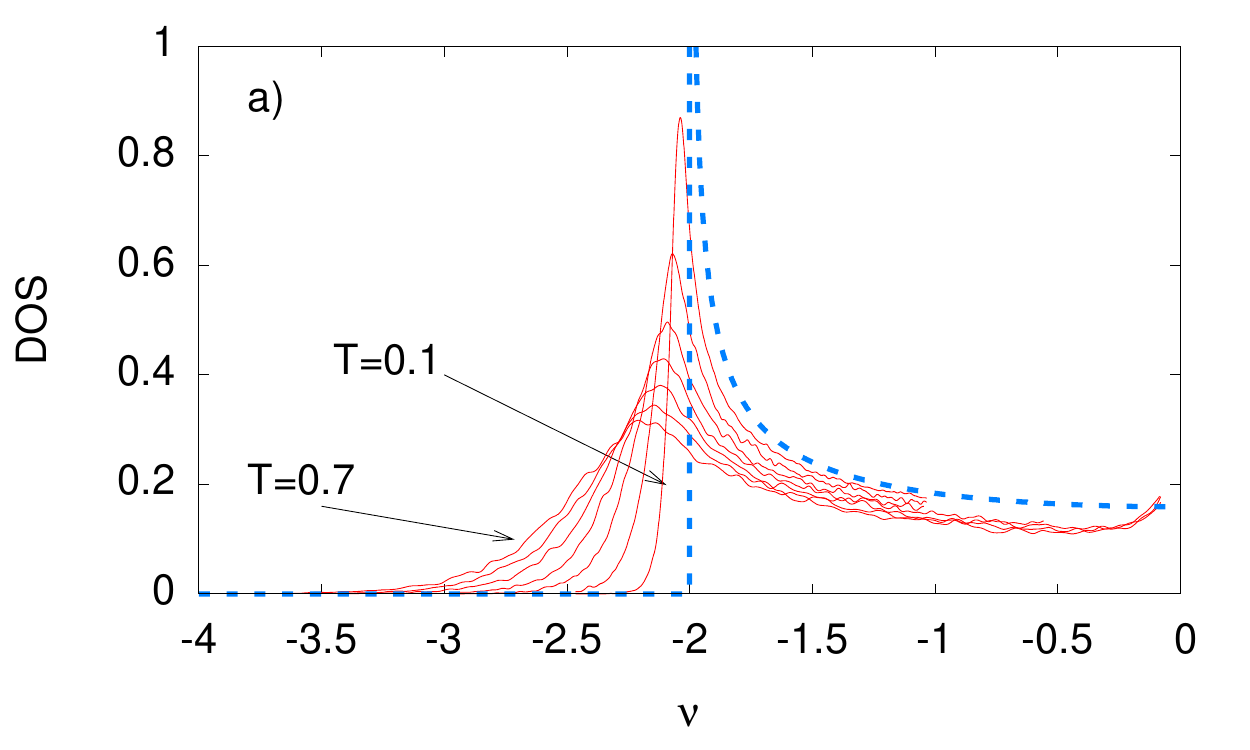}
\includegraphics[width=8cm]{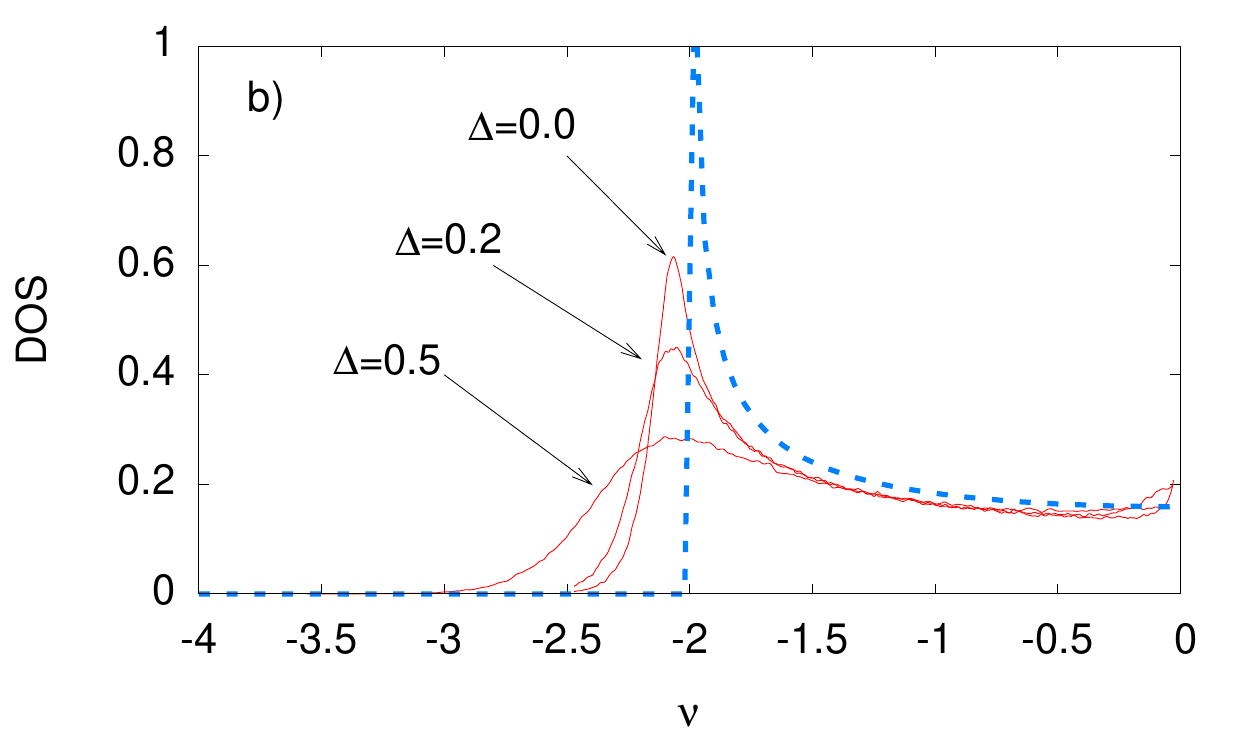}
\caption{(a) Density of states in  a pure crystal  with intrinsic
  inter-molecular  disorder and $\Delta=0$. 
The different curves correspond to increasing  temperatures from 
$T=0.1t_0$ to $T=0.7t_0$. The dashed line is the DOS of the 
perfect crystal in the absence of thermal disorder 
($\lambda=0$). (b) DOS at $T=0.2t_0$ for
several values of the extrinsic disorder $\Delta$. }
\label{fig:DOS}
\end{figure*}

Based on Eq. (\ref{eq:mobzero}), 
the electron mobility in a non-degenerate semiconductor 
arises from a weighted  average 
of the energy-resolved diffusivity
$D(\nu)$ via the thermal population of electronic
states. The latter is 
measured by the thermally weighted DOS, 
\begin{equation}
\label{eq:wDOS}
W(\nu)=\frac{\rho(\nu)e^{-\beta
  \nu}}{\int \rho(\nu)e^{-\beta
  \nu}},
\end{equation}
which represents the normalized probability of occupation of the
states at energy $\nu$. The functions $\rho(\nu)$, $D(\nu)$ and
$W(\nu)$ are analyzed below.

Fig. \ref{fig:DOS} illustrates the evolution of the 
DOS $\rho(\nu)$ as a function of increasing thermal disorder
 in a pure sample (a) and upon increasing extrinsic 
disorder (b). The DOS of a perfectly 
ordered crystal is shown for reference (dashed).
In both cases, the Van-Hove singularity marking the  edge of the
one-dimensional band at
$\nu=-2t_0$ is 
rounded off and shifts deeper in energy, indicating an increase of the
effective bandwidth. \cite{HoHu,orgarpes11,Coropceanu12}   
In addition, tails are generated beyond the range of band states.
Both the bandwidth increase and the extension of band tails
 are  controlled  by the amount of disorder. This can be
quantified through the variance
  $s=\sqrt{8\lambda T t_0}$ of the thermal fluctuations of the inter-molecular
  transfer intergrals in the intrinsic case 
\cite{reconcile09} (Fig.\ref{fig:DOS}a), and by the spread $\Delta$ of
molecular energy levels in the
extrinsic case \cite{HoHu} (Fig.\ref{fig:DOS}b).

\begin{figure*}
\centering
\includegraphics[width=8cm]{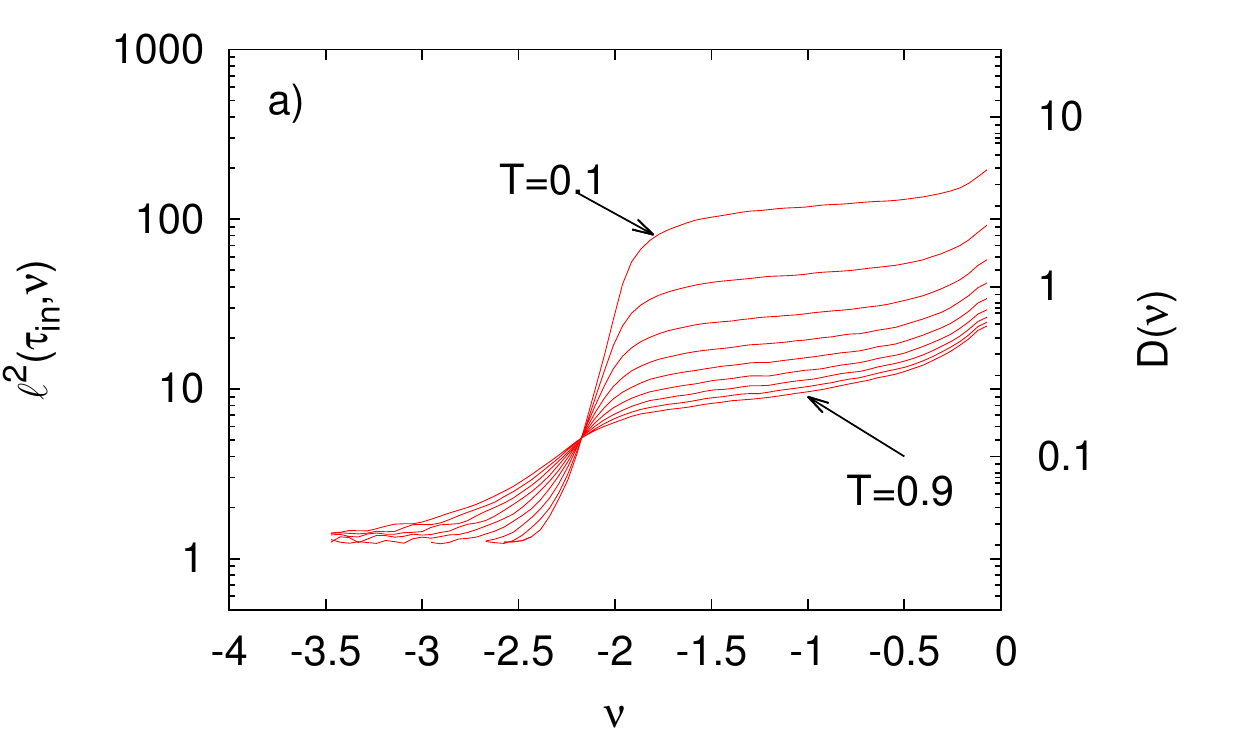}
\includegraphics[width=8cm]{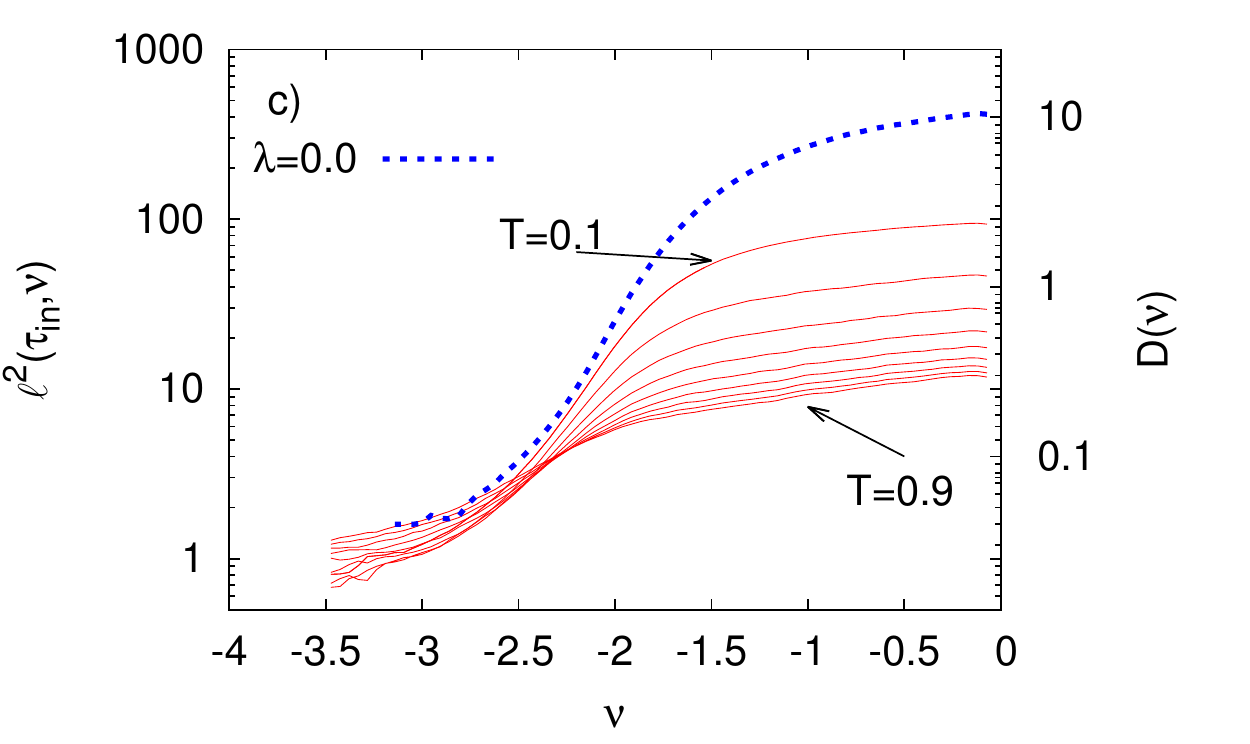}
\includegraphics[width=8cm]{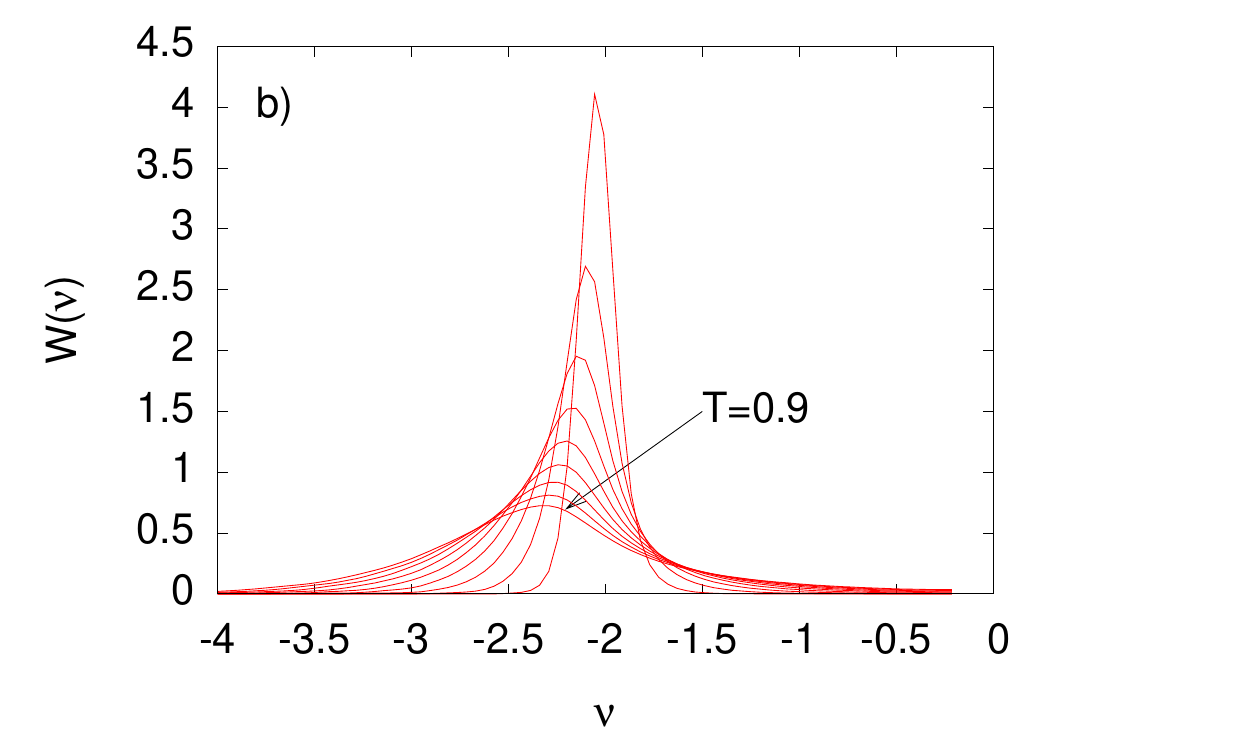}
\includegraphics[width=8cm]{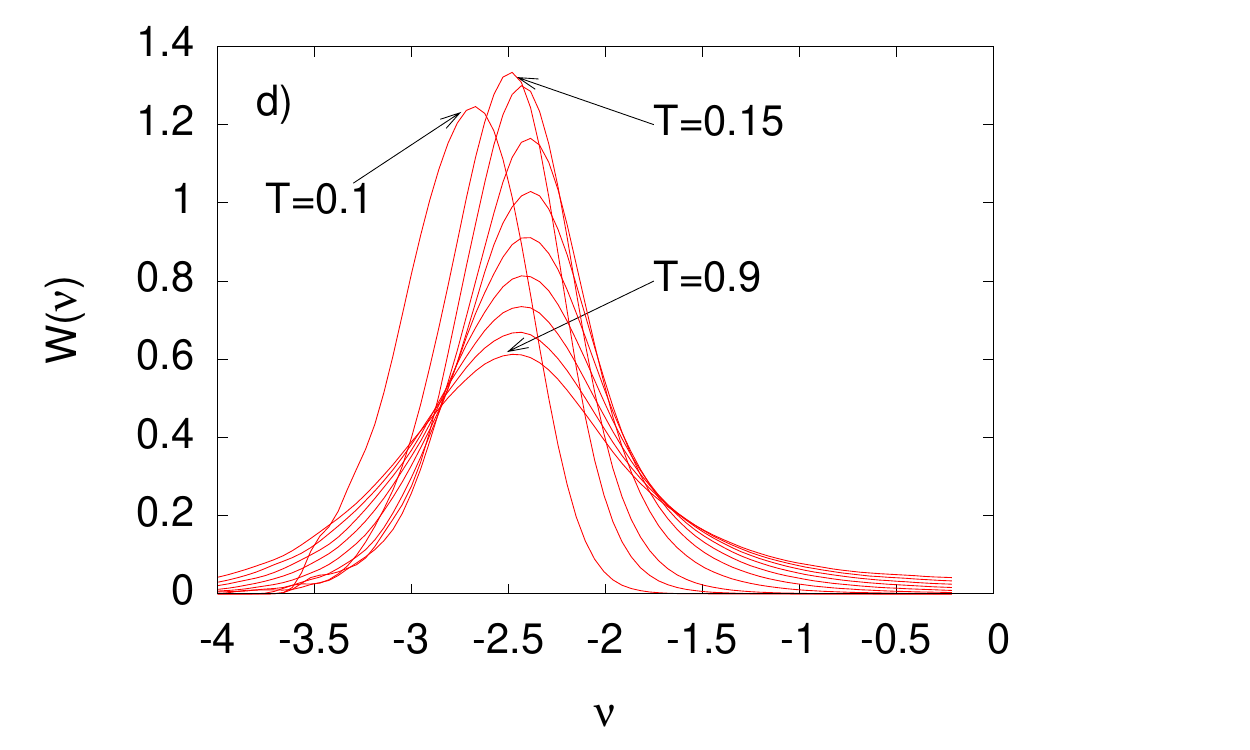}
\caption{(a) Energy-resolved transient localization length squared $\ell^2
  (\tau_{in},\nu)$ (left axis) and diffusivity $D(\nu)$ (right axis) 
   in a pure OSC, for increasing temperatures from 
  $T=0.1t_0$ to $T=0.9t_0$. Parameters are the same as in the
  preceding figures. The length unit is the lattice spacing $a$. 
  (b) Weighted DOS [see Eq. (\ref{eq:wDOS})] at the same temperatures. 
  (c) and (d) show the same quantities in the presence of extrinsic disorder 
  ($\Delta=0.5$). In panel (d) the temperature 
  $T=0.15t_0$ has been  added to highlight the crossover from extrinsic to 
intrinsic transport.} 
\label{fig:ellBubble}
\end{figure*}

The diffusivity $D(\nu)$ obtained through the RTA in the pure case 
 ($\Delta=0$) is shown 
 in Fig. \ref{fig:ellBubble}a (right axis scale). The diffusivity  is  directly related,
through  Eq. (\ref{eq:RTAbub}), to the square of the energy-resolved 
transient localization length, $\ell^2(\tau_{in},\nu)$, which is also shown on
the same figure on the left axis. Analogous estimates for the localization length in the
static disorder limit $\tau_{in}\to \infty$ can be found in Refs. \onlinecite{reconcile09,Chang11}.
The comparison  with the DOS of  Fig. \ref{fig:DOS}a 
allows us to identify two distinct regions in the electronic
spectrum, separated by a crossover region of width $\simeq s$ around
the band edge. States 
located in the band region
have a large diffusivity, that is strongly suppressed upon
increasing the thermal disorder.
Tail states induced by 
disorder below the band edge instead
 have a much lower diffusivity as a consequence of their more localized
character. 
The diffusivity of tail states 
is essentially temperature independent and corresponds in our
one-dimensional model to a minimum
localization length of approximately one lattice spacing,
$\ell(\tau_{in},\nu)\approx a$.
We note that the existence of two distinct characteristic values of
the localization length is in agreement with recent ESR measurements
performed on pentacene transistors.\cite{Marumoto,Matsui,Mishchenko}

The relative importance of band and tail states in the transport mechanism is 
determined by  the weighting function $W(\nu)$,
which is shown in Fig. \ref{fig:ellBubble}b.
At temperatures $T\lesssim 0.5t_0$, which includes the experimentally
accessible range, the function $W(\nu)$ is
peaked right in the 
crossover range that separates the band and tail states
(cf. Fig. \ref{fig:DOS}),  with a 
sizable overlap on both sides. As expected for two conduction channels in
parallel,  the electronic transport in this case
 is dominated by the channel whose diffusivity is largest, i.e. the band
 states. Correspondingly, the
 temperature dependence of the mobility is governed by the
 suppression of the diffusivity in the band range, that is illustrated in 
Fig. \ref{fig:ellBubble}a. Upon increasing the 
temperature, the weighting function progressively broadens and
 shifts towards the tail states. These eventually become
 the dominant transport channel, leading to the mobility saturation observed in
 Fig. \ref{fig:mobility}a.

\begin{figure}
\centering
\includegraphics[width=8cm]{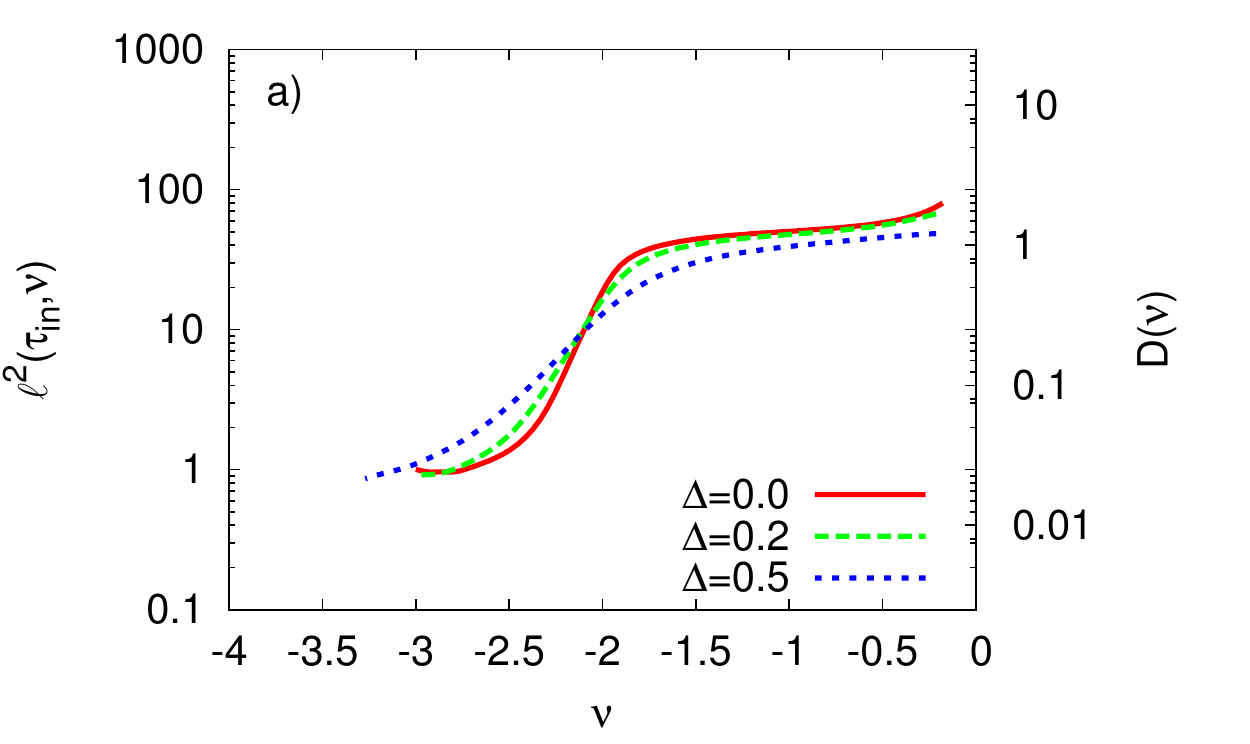}
\includegraphics[width=8cm]{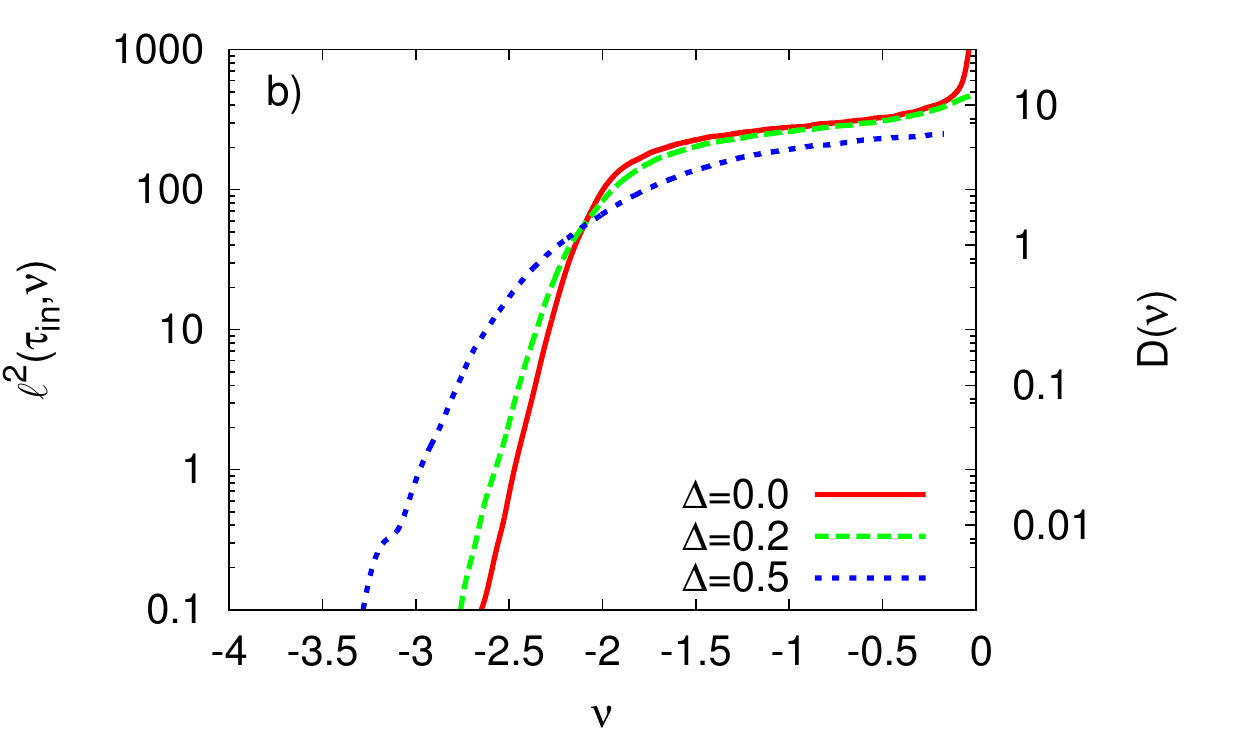}
\caption{ 
 Energy resolved localization length squared $\ell^2 (\tau_{in},\nu)$ 
within the RTA (a) and  Kubo bubble approximations (b). In both cases we
 take $1/\tau_{in}=\omega_0=0.05t_0$} 
\label{fig:ell-vs-Delta}
\end{figure}

The energy-resolved diffusivity and transient localization length 
obtained in the presence of extrinsic disorder ($\Delta=0.5t_0$) are
illustrated in Fig. \ref{fig:ellBubble}c.
The main difference with  the pure case shown in 
Fig. \ref{fig:ellBubble}a is that 
the crossover region separating tail and band states is now broadened 
by an amount $\propto \Delta$. Nevertheless, the typical values of the 
diffusivity both in the band region and in the tails remain close to their
 intrinsic values, provided that the temperature is not too low.  
This is more clearly seen in
 Fig. \ref{fig:ell-vs-Delta}a, which shows the RTA diffusivity 
at increasing values of $\Delta$ for fixed $T=0.2t_0$.
At lower temperatures ($T\le 0.1t_0$) the extrinsic disorder eventually 
becomes dominant and the results tend to recover 
those obtained in the absence of intrinsic electron-vibration coupling
($\lambda=0$, dashed curve in Fig. \ref{fig:ellBubble}c).

Since for $\Delta = 0.5t_0$ 
the diffusivity is a monotonically decreasing function of
$T$ for all states, the origin of the activated behavior of the mobility 
observed  at low temperatures in Fig. \ref{fig:mobility}a
 has to
be sought elsewhere, i.e. 
in the weighting function $W(\nu)$. As illustrated in 
Fig. \ref{fig:ellBubble}d,  
the behavior of $W(\nu)$ in the presence of extrinsic
 disorder is opposite to that of the
pure case, Fig. \ref{fig:ellBubble}b:  at  low temperature (here $T=0.1t_0$)
 the peak of the weighting function is located deep in
the tail states and it moves towards the band
 upon increasing the temperature. Tail states with a low diffusivity therefore
 dominate the transport mechanism at low temperature, while
the intrinsic regime is progressively recovered upon increasing the 
temperature. The
 crossover between these two regimes is signaled by a maximum in the
 mobility of Fig. \ref{fig:mobility}a at a temperature that we denote $T^*$. 
By comparing the variances of intrinsic and extrinsic disorder given
 at the beginning of this Section, i.e. setting $s\simeq \Delta$, 
 we obtain the following estimate for
 the crossover temperature: $T^*\simeq
 \Delta^2/(8\lambda t_0)$. Taking $\Delta=0.5t_0$ and $\lambda=0.17$
  gives $T^*=0.18t_0$, in good  agreement 
with the data of Fig. \ref{fig:mobility}a. The predicted crossover
temperature for
$\Delta=0.2t_0$ is $T^*=0.03t_0$, outside the studied range.

The Kubo bubble approximation yields qualitatively similar results for
the temperature and $\Delta$ dependence of the diffusivity $D(\nu)$ (of
course the DOS and weighting function $W(\nu)$ are exactly the
same, as they are obtained in the common static limit). 
The main differences between the two methods  
are quantitative and  arise
from the inclusion or not of backscattering effects, 
i.e. they are indicative of the 
relevance of vertex corrections in the disordered system under study. 
As shown in  Fig. \ref{fig:ell-vs-Delta},
the diffusivity of band states in the Kubo bubble approximation 
 is larger than in the RTA,  corresponding to the fact that for band states
 the Kubo bubble essentially recovers the Boltzmann
 transport theory\cite{reconcile09} where quantum localization phenomena are
 absent.  
This leads to 
larger values of the intrinsic mobility than in the RTA, 
as seen in Fig. \ref{fig:mobility}b. 
Concerning tail states, the opposite is true. 
In the RTA the localization length and hence the diffusivity appear 
to be bound from below,
which  provides a lower bound to the mobility 
at low temperatures: taking $\ell(\tau_{in},\nu)> a$ from
Fig. \ref{fig:ell-vs-Delta}a yields $\mu_e> ea^2/(2\tau_{in} k_B
T)$.  
This lower bound is absent in the Kubo bubble approximation,
where $D(\nu)$ vanishes asymptotically
for negative energies, being itself proportional to the DOS. 
It can actually  be shown  \cite{Coehoorn} 
that a  behavior 
of the form $\mu_e\propto e^{-(\Delta/2T)^2}$ is obtained from the
Kubo bubble in the
limit of strong disorder, $\Delta\gg t_0$.
As a result 
the thermal activation  at low temperatures is more pronounced in
the Kubo bubble approximation than in the RTA.
The two effects discussed here are at the origin of the
 much sharper crossover from 
extrinsic to intrinsic transport obtained in the Kubo bubble compared
to the RTA, cf. Fig. \ref{fig:mobility}.

\section{Density  dependence of the mobility}
\label{sec:density}

\begin{figure}
\centering
\includegraphics[width=8cm]{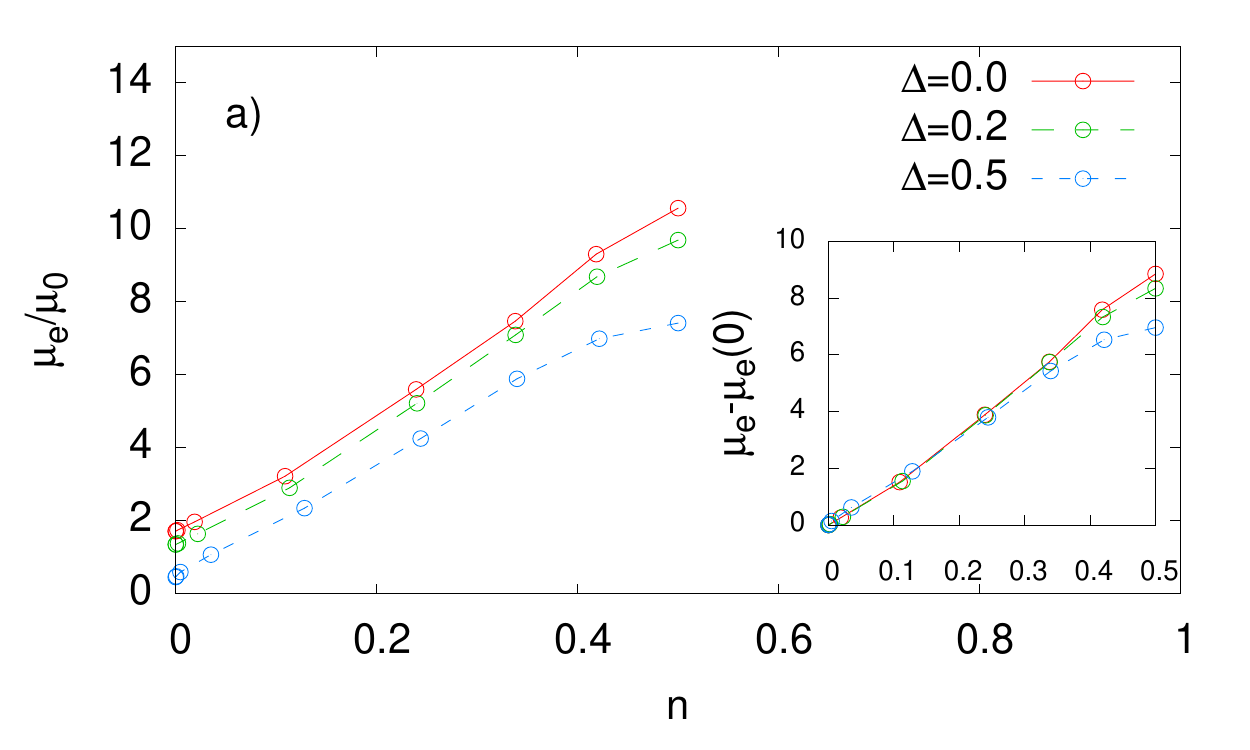}
\includegraphics[width=8cm]{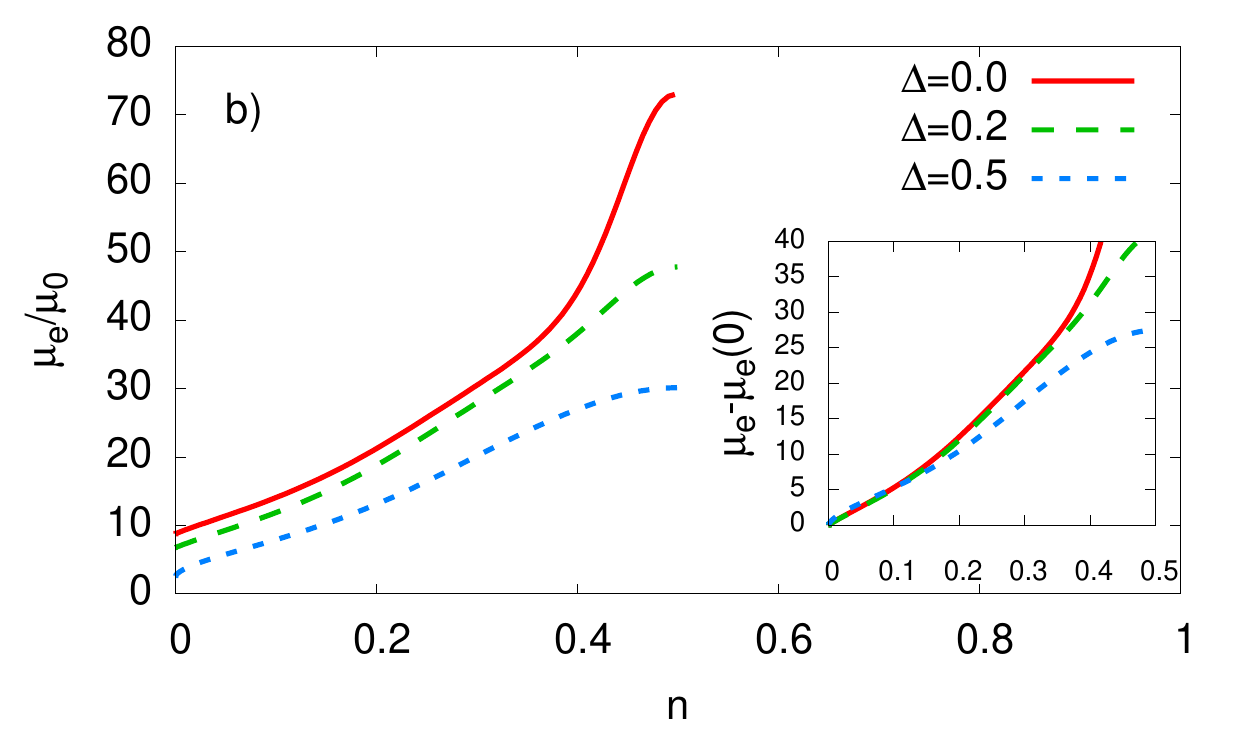}
\caption{Mobility as a function of carrier concentration per molecule
  within (a) the RTA and (b) Kubo bubble approximation, for $T=0.2t_0$. 
 The inset shows the variation of mobility with respect to its zero
 density limit. }  
\label{fig:mobility-vs-dens}
\end{figure}

The  mobility obtained through Eq. (\ref{eq:mobdens}) 
at finite electron concentration is shown in
Fig. \ref{fig:mobility-vs-dens}. The parameters are the same as in
Fig. \ref{fig:CpAll}, i.e. $\lambda=0.17$, $T=0.2t_0=300K$.  
In both the  RTA, Fig. \ref{fig:mobility-vs-dens}a, and Kubo bubble
approximation, Fig. \ref{fig:mobility-vs-dens}b, 
 we find a steady increase of the mobility with
increasing density. 
This behavior can be understood as arising from a progressive  filling of tail
states, which occurs through a shift of the chemical potential 
towards the band region  as the 
electron liquid becomes degenerate. This allows
states with a higher diffusivity to be populated, 
via a shift of the the factor $-\partial
f/\partial \nu$  in Eq. (\ref{eq:mobdens}).
This argument is only qualitatively correct
however, because it neglects the fact that 
the diffusivity $D(\nu)$ itself depends on the density,
which is instead 
correctly incuded in the results of Fig. 
\ref{fig:mobility-vs-dens}.
Based on  the same argument, increasing the
density of carriers will shift the crossover between the extrinsic and
intrinsic regimes to lower temperatures. 
Such depinning effect 
 can be achieved in OFETs through the application of 
 a strong enough gate electric
field.

We see from Fig. \ref{fig:mobility-vs-dens} that  
the curves describing the density dependence of the mobility for different
degrees of extrinsic disorder are essentially parallel. 
The quantity
$\mu_e-\mu_e(n=0)$, 
where the $n=0$ limiting value has been subtracted, is shown in the
inset. Interestingly, it appears to be insensitive to the presence of extrinsic
sources of disorder:  the curves show a clear collapse   
for different values of the static
disorder, that persists up to fairly large carrier concentrations.
This indicates that at the considered temperature, $T=0.2t_0=300K$,
the observed increase of mobility with density comes
entirely from populating carriers with a strong band character 
 (cf. Fig. \ref{fig:ellBubble}). 
The quantity $\mu_e-\mu_e(n=0)$  could therefore be used to
measure the mobility of band carriers even in samples
with a sizable degree of disorder. 

\begin{figure}
\centering
\includegraphics[width=8cm]{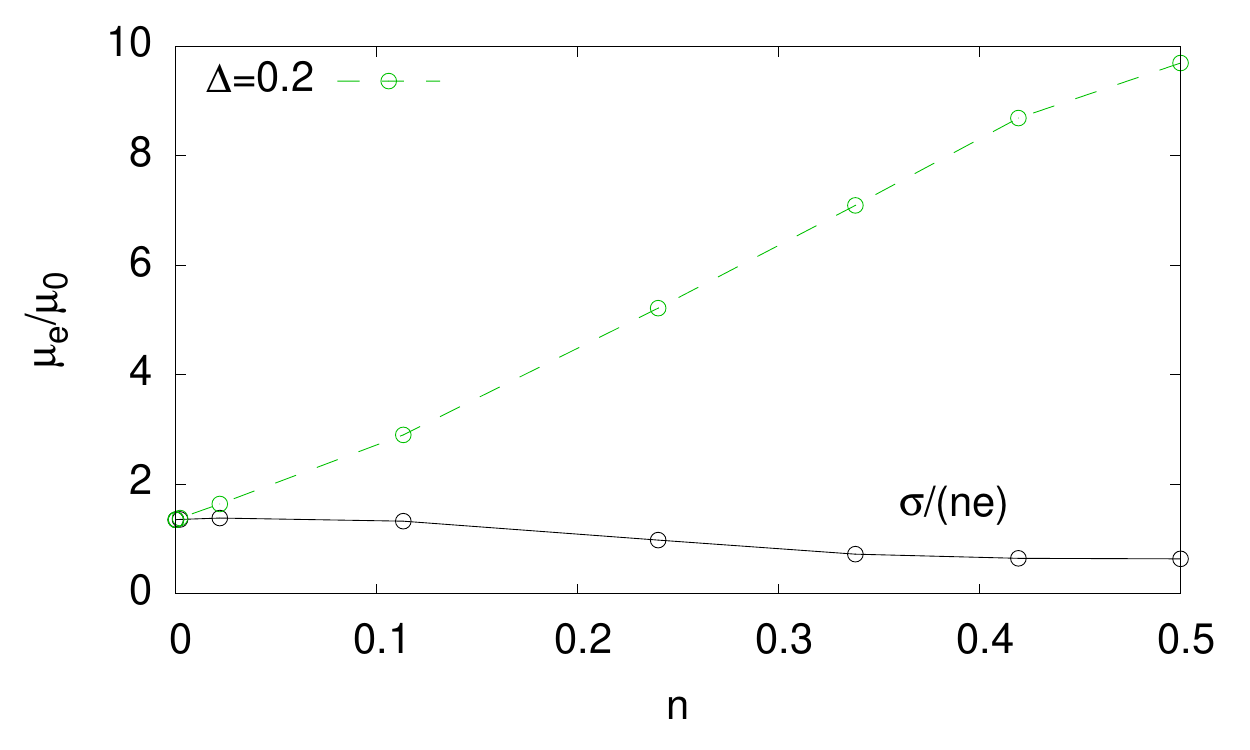}
\caption{Mobility as a function of carrier concentration per molecule, 
obtained from the RTA 
for $T=0.2t_0$ and $\Delta=0.2t$ (dashed line, same as Fig. 6a). 
The full line is obtained by replacing
the definition Eq. (\ref{eq:mobfin}) by the expression
$\mu_e=\sigma/(ne)$.}  
\label{fig:compare}
\end{figure}

Finally, because large carrier concentrations 
are now customarily obtained in OFETs 
via liquid gating \cite{Frisbie,Iwasa}, 
the correct definition of the 
mobility, Eq. (\ref{eq:mobfin}), should be used to analyze experiments, 
instead of the usual expression 
$\sigma/(ne)$ that only holds in the limit of vanishing density. 
To illustrate this point, we compare the results obtained with the two 
definitions in Fig. \ref{fig:compare}. Use of the low-density expression 
(black, full line) leads to 
an erroneous result, as it  incorrectly predicts a reduction of the mobility 
upon  increasing the density. To understand this result,
we observe that for degenerate carriers, the susceptibility Eq. (\ref{eq:compr})
is given by the DOS at the chemical potential, i.e. 
$\partial n/\partial \mu \simeq \rho(\mu)$. Using Eq. (\ref{eq:mobfin})
we can then rewrite $\sigma/(ne)\simeq \mu_e  k_BT \rho(\mu)  /n$, 
which is itself proportional to the DOS (cf. Fig. 3b). In the present model an increase 
of the density implies a reduction of the DOS at the chemical potential,
explaining the behavior observed in Fig. \ref{fig:compare}.


\section{Conclusions}
\label{sec:conclusions}

Based on a theoretical formalism that relates the Kubo formula for the
conductivity 
to the time-resolved diffusivity of electronic states, we
have analyzed the electronic transport mechanism in a model that
accounts for several key ingredients relevant to organic semiconductors: 
the existence of narrow electronic bands, the
dynamical  disorder arising from the thermal vibrations of the
molecules and the presence of extrinsic sources of disorder that are
 unavoidable in real samples and devices. 

The presence of strong dynamical 
disorder is intrinsic to organic semiconductors and invalidates
the usual semiclassical treatments of electronic transport that apply
to inorganic semiconductors, calling for a theoretical
 approach that is able to treat quantum localization corrections in
a controlled way. This is achieved here through a relaxation time
approximation (RTA) that relates directly the carrier diffusivity to
the localization properties of the electronic states. 
Within this theoretical scenario,
the deviations from semiclassical transport are
understood as arising from a {\it transient localization} of
electrons, that takes place before the onset of a true diffusive
behavior at long times. This  phenomenon
appears to be a characteristic feature 
of organic semiconductors, where the typical  timescale of inter-molecular
vibrations is longer than the elastic scattering time.
The transient localization scenario is supported by numerical simulations on the
time-dependent diffusivity \cite{RTA11} and by optical conductivity
measurements in Rubrene OFETs \cite{Li,Fischer}.

Based on the present theory, 
the intrinsic transport mechanism 
in clean organic semiconductors  
is explained as the diffusive
spread of localized wavefunctions rather than the scattering of
delocalized waves by phonons and disorder. 
A power-law decay with temperature is predicted for 
the intrinsic mobility, which results from
the reduction of the transient localization length as the
thermal disorder increases. Our results suggest that the intrinsic mobility
of organic semiconductors could be improved by tailoring crystal structures 
with stiffer inter-molecular bonds, as this would reduce the impact of thermal 
disorder on the charge transport.

The inclusion of extrinsic disorder causes a crossover from the
intrinsic power-law behavior, persisting at high temperature, towards a
thermally activated behavior induced by carrier trapping 
at low temperature. 
Increasing the electron concentration
induces a depinning from trapped states,
leading to  an increase of the mobility 
and a progressive suppression 
of the thermally activated regime. 
Our results for the concentration  dependence of the mobility
 generalize the findings obtained  in the classical hopping
 limit $t_0 \ll \Delta$  \cite{Bassler,Coehoorn}
to the high mobility organic FETs of present interest, where
the existence of electronic bands requires a
quantum treatment of electron motion.

From a more general viewpoint, 
the present work demonstrates that 
the conductive properties of both
pure and disordered
organic semiconductors can be efficiently understood within a unified
framework, by addressing the interplay
between mobile  states in the band region and strongly localized states
in the band tails. 
The present results confirm and extend the considerations of 
Ref. \onlinecite{reconcile09} by allowing for a proper inclusion of
quantum localization phenomena. 
Interestingly, 
the relationship between the energy-resolved properties of electronic 
states and the resulting mobility, that we have exploited here,   
could also be  generalized to study how the intrinsic
polarizability of the organic crystals affects the transport
characteristics, as was recently proposed in  
Ref. \onlinecite{Minder}.

\appendix

\section{Detailed balance and spectral representation}
\label{app:detbal}
In this Appendix we set $e=1$ in addition to $\hbar=k_B=1$. 
By introducing 
the Laplace transform of the the retarded current-current correlation
function
$C_+(t)$ 
\begin{equation}
C_+(p) = \int_0^\infty dt e^{-pt} C_+(t),
\label{eq:LapCp}
\end{equation}
we can write the mean square displacement 
defined in Eq. (\ref{eq:defL2}) as
\begin{equation}
\label{eq:L2pCp}
L^2(p)=\frac{C_+(p)}{p}.
\end{equation}
Expressing $C_+(t)$ in terms of its Fourier transform $C_+(t)=\int \frac{d\omega}{2\pi} e^{-i\omega t} C_+(\omega)$
and using the reality of $C_+(t)$, which implies 
that $ReC_+(\omega)$ and $Im
C_+(\omega)$ are respectively an even and an odd 
function of $\omega$, we have
\begin{equation}
\label{eq:CpCw}
C_+(p) = \int_0^\infty \frac{d\omega}{\pi} \frac{p ReC_+(\omega) + \omega ImC_+(\omega)}{p^2+\omega^2}. 
\end{equation}
This equation can also be obtained using the analiticity of
$C_+(\omega)$ in the complex upper half-plane with the help of
Cauchy's residue theorem for complex integration.
The two terms in Eq. (\ref{eq:CpCw}), respectively proportional to the
real and imaginary part of $C^+$, 
bring the same contribution to the integral 
as can be checked via the Lehman
representation of the correlation function 
\begin{equation}
\label{eq:LehmanCt}
C_+(t)= \frac{1}{Z} \sum_{n,m} e^{-\beta E_n} |<n|J|m>|^2 2\cos (\omega_{n,m}t),
\end{equation}
where  $|n>$ and $E_n$ are respectively the
eigenvectors and eigenvalues of the Hamiltonian, that are supposed to
be known, and $\omega_{n,m}=E_n-E_m$. 
The Fourier transform of Eq. (\ref{eq:LehmanCt}) reads
\begin{widetext}
\begin{eqnarray}
Re C_+(\omega)&=&\frac{\pi}{Z}\sum_{n,m} e^{-\beta E_n} |<n|J|m>|^2
\left[ \delta(\omega-\omega_{n,m})+\delta(\omega+\omega_{n,m})\right] \label{eq:ReC}  \\
Im C_+(\omega)&=&\frac{1}{Z}\sum_{n,m} e^{-\beta E_n} |<n|J|m>|^2 \frac{2\omega}{\omega^2-\omega^2_{n,m}} \label{eq:ImC}
\end{eqnarray}
\end{widetext}

Using the result
\begin{equation}
P \int_0^\infty \frac{d\omega}{\pi} \frac{2 \omega^2}{p^2+\omega^2}\frac{1}{\omega^2-\omega^2_{n,m}}=\frac{p}{p^2+\omega^2_{n,m}}
\end{equation}
we arrive at
\begin{equation}
\label{eq:CpCw2}
C_+(p) = \int_0^\infty \frac{d\omega}{\pi} \frac{ 2p}{p^2+\omega^2}  ReC_+(\omega). 
\end{equation}
From this equation we can prove the limit used to obtain the
Einstein's relation  
Eq. (\ref{eq:Einstein}),
 observing that as $p\rightarrow 0^+$ $\frac{1}{\pi} \frac{p}{p^2+\omega^2}\rightarrow \delta(\omega)$ then $\lim_{p\to 0}
C_+(p)=\lim_{\omega\to 0} C_+(\omega)/2$.

Eqs. (\ref{eq:Kubocompact}) and (\ref{eq:Laplace}) are two relations
which are not independent since the functions $C_+$ and $C_-$ 
are related by the detailed balance condition.
The Fourier transforms of the retarded correlation functions
 $C_\pm(t)$ can be expressed as
$C_\pm(\omega) = C_>(\omega)\pm C_<(\omega)$ where $C_>(\omega),C_<(\omega)$ are 
the Fourier transforms
\begin{eqnarray}
\label{eq:auxiliary1}
C_>(\omega) &=& \int_{0}^{+\infty} dt e^{i(\omega+i\delta)t} <\hat J(t) \hat J(0)>\\
\label{eq:auxiliary2}
C_<(\omega) &=& \int_{0}^{+\infty} dt e^{i(\omega+i\delta)t} <\hat J(0) \hat J(t)>
\end{eqnarray}
and $\delta$ an infinitesimal positive quantity.
The detailed balance condition reads 
\cite{Baym} 
$Re
C_<(\omega)=e^{-\beta\omega} Re C_>(\omega)$, which leads to
\begin{equation}
\label{eq:detbalapp}
Re C_-(\omega)=\tanh \left(\frac {\beta \omega}{2}\right) Re C_+(\omega).
\end{equation}

Applying the detailed balance Eq. (\ref{eq:detbalapp}) to Eq. (\ref{eq:CpCw}) 
yields
\begin{equation}
\label{eq:CpCm}
C_+(p) = \int_0^\infty \frac{d\omega}{\pi} \frac{ 2p\;Re C_-(\omega)
}{(p^2+\omega^2) \tanh (\beta \omega/2)} 
\end{equation}
and using the Kubo formula Eq. (\ref{eq:Kubocompact}) 
with the definition Eq. (\ref{eq:L2pCp}) we arrive at
 \begin{equation}
\label{eq:L2sigma}
L^2(p) = \int_0^\infty \frac{d\omega}{\pi} \frac{ 2\omega \sigma(\omega) }{(p^2+\omega^2) \tanh (\beta \omega/2)}
\end{equation}
which is Eq. (\ref{eq:L2omega}) of the paper.

Similarly to Eq. (\ref{eq:ReC}), the Fourier transform of the anticommutator correlation function $C_-(\omega)$ 
can be expressed through its Lehman representation as
\begin{widetext}
\begin{equation}
Re C_-(\omega)=\frac{\pi}{Z}\sum_{n,m} e^{-\beta E_n} |<n|J|m>|^2 \left[ \delta(\omega+\omega_{n,m})-\delta(\omega-\omega_{n,m})\right].
 \label{eq:ReCm}
\end{equation}
The two terms in the sum can be rearranged as 
\begin{equation}
Re C_-(\omega)= \pi (1-e^{-\beta\omega})\frac{1}{Z}\left \{ \sum_{n,m}e^{-\beta E_n}
 |<n|\hat J|m>|^2  \delta(\omega+E_n-E_m) \right \}.
\label{eq:Lehman2}
\end{equation}

We now  consider single-particle Hamiltonians for which 
the number eigenstates 
$|\{n_\alpha\}>$ are such that $H|\{n_\alpha\}>=(\sum_\beta n_\beta \epsilon_\beta)|\{n_\alpha\}>$.
Once expressed using this basis Eq. (\ref{eq:Lehman2}) takes the form (in the grand canonical ensemble)
\begin{equation}
Re C_-(\omega) = \pi (1-e^{-\beta\omega})\frac{1}{Z}\left \{ \sum_{\{n\} \{m\}} \Pi_\alpha e^{-\beta (\epsilon_\alpha-\mu)n_\alpha}|<\{n_\alpha\}|\hat J|\{m_\alpha\}>|^2
\delta[\omega+\sum_\alpha \epsilon_\alpha (n_\alpha-m_\alpha)] 
 \right \}.
\label{eq:Lehman3}
\end{equation}
Since $\hat J$ is a single-particle operator, the individual
eigenvalues obey 
$n_\alpha-m_\alpha=p_\alpha$ with $p_\alpha=-1,0,1$.
Thus in the sum appearing in the $\delta$ function reduces to $\epsilon_\alpha-\epsilon_\beta$ where we have 
$p_\gamma=1$ when $\gamma=\alpha$,
$p_\gamma=-1$ when $\gamma=\beta$ and $p_\gamma=0$ elsewhere.
The matrix element of $\hat J$  reads in this case
\begin{equation}
|<\{n_\alpha\}|\hat J|\{m_\alpha\}>|^2 = \sum_{\alpha,\beta} |<\alpha|\hat J|\beta>|^2 n_\alpha (1-n_\beta)
\end{equation}
where $|\alpha>$ are single-particle states. The grand canonical averages appearing in Eq. (\ref{eq:Lehman3}) can be performed leading to
\begin{equation}
Re C_-(\omega) = \pi (1-e^{-\beta\omega})
\sum_{\alpha,\beta}
|<\alpha|\hat J|\beta>|^2<n_\alpha> (1-<n_\beta>)\delta(\omega+\epsilon_\alpha-\epsilon_\beta) .
\label{eq:Lehman4}
\end{equation}
where we have made use of the vanishing of the diagonal elements of $\hat J$.
We can introduce a dummy integration variable by writing
\begin{equation}
Re C_-(\omega) = \pi (1-e^{-\beta\omega}) \int_{-\infty}^{\infty} d\nu f(\nu) [1-f(\omega+\nu)]
\sum_{\alpha,\beta}
|<\alpha|\hat J|\beta>|^2\delta(\omega+\nu-\epsilon_\beta)\delta(\nu-\epsilon_\alpha).
\label{eq:Lehman5}
\end{equation}
where $f(\nu)$ is the Fermi function.
 Identifying the diagonal part of the spectral operator
$\delta(\nu-\epsilon_\alpha)=-\frac{1}{\pi}Im<\alpha|(\nu -\hat H)^{-1}|\alpha>$ and taking into account that 
$(1-e^{-\beta\omega})f(\nu)(1-f(\omega+\nu))=f(\nu)-f(\omega+\nu)$ we finally arrive at
\begin{equation}
Re C_-(\omega) = \pi (1-e^{-\beta\omega}) \int_{-\infty}^{\infty} d\nu f(\nu) (1-f(\omega+\nu))
tr \left [  \hat \rho(\nu) \hat J \hat \rho(\omega + \nu) \hat J\right ].
\label{eq:Kuboexpandapp}
\end{equation}

Using Eq. (\ref{eq:Kuboexpandapp}) and Eq. (\ref{eq:CpCm}) we obtain:
\begin{equation}
\label{eq:L2pE}
L^2(p)=\frac{2}{e^2} \int_{-\infty}^{\infty}
 d\nu \int_0^\infty d\omega \frac{1}{\omega^2+p^2}  \frac{\left[ f(\nu)-f(\omega+\nu)\right]}{\tanh
  (\beta \omega/2)} \ tr  [\hat \rho(\nu) \hat J \hat \rho(\nu+\omega) \hat J]
\end{equation}
Taking the low density limit, $\beta\mu\to -\infty$, we obtain
\begin{equation}
\label{eq:L2pElowdens}
\frac{L^2(p)}{n}=\frac{2}{e^2 Z} \int_{-\infty}^{\infty}
 d\nu e^{-\beta \nu}\int_0^\infty d\omega \frac{{1+e^{-\beta \omega}}}{\omega^2+p^2}  
   \ tr  [\hat \rho(\nu) \hat J \hat \rho(\nu+\omega) \hat J]
\end{equation}
where the normalization factor is $Z={\int d\nu e^{-\beta \nu}\rho(\nu)}$.
\end{widetext}

\section{Optical conductivity sum-rules}
\label{app:sum-rules}

Eq. (\ref{eq:L2omega})
allows us to derive exact relationships between
 the asymptotic expansion of $C_+(p)$ [or
equivalently $L^2(p)$] to certain integrals of 
the optical conductivity. 
Using the definition Eq. (\ref{eq:L2pCp}) we write
Eq. (\ref{eq:L2sigma}) 
 as
\begin{equation}
  \label{eq:Cpomega}
  C_+(p)=\int_0^{\infty} \frac{d\omega}{\pi}
  \frac{2p\omega}{\omega^2+p^2}\frac{\sigma(\omega)}{\tanh(\beta
    \omega/2)}.  
\end{equation}
Taking the formal expansion in powers of $1/p$ we get
\begin{equation}
  C_+(p)=\sum_{n=0}^{\infty} \frac{(-1)^n}{p^{2n+1}}S_n 
\end{equation}
where 
\begin{equation}
S_n = \int_0^{\infty} \frac{d\omega}{\pi}
  \frac{2\omega^{2n+1}}{\tanh(\beta\omega/2)} {\sigma(\omega)}.
\end{equation}
Taking into account the definition of the Laplace transform we obtain
\begin{equation}
  C_+(p)=\sum_{n=0}^{\infty} \frac{2n+1}{p^{2n+1}} \frac{d^n C_+(t)}{d
    t^n}|_{t=0}.
\end{equation}
The leading term of the asymptotic expansion ($n=0$) gives $C_+(p)\simeq C_+(t=0)/p$ where
$C_+(t=0)=<J^2>$.

Equating the coefficients of the expansions we get
\begin{equation}
  \frac{d^n C_+(t)}{d t^n}|_{t=0} = (-1)^n S_n
\end{equation}
which reads for $n=0$
\begin{equation}
  C_+(t=0) = \int_0^{\infty} \frac{d\omega}{\pi}
  \frac{2\omega}{\tanh(\beta\omega/2)} {\sigma(\omega)}.
\end{equation}
Setting $p=1/\tau$ as in Fig. 1 we have that in the short time
limit $C_+(\tau)=C_+(t=0)\tau=<J^2> \tau$. While the RTA obeys this
sum rule because the correlation function $C^+$ is exact at short
times, the Kubo bubble approximation does not, which results in the
slight discrepancy observed 
  in Fig. 1 in the ballistic regime. This  can be related via Eq. (B6) 
to the different behaviour obtained for 
$\sigma(\omega)$ in the two approximations and  points 
to the relevance of vertex corrections at all frequencies.

\section{Boltzmann theory from the RTA}
\label{sec:BoltzmannRTA}
 
The quantity $\ell(p,\nu)^2$ is evaluated in practice from 
\begin{equation}
\ell^2(p,\nu) = \frac{2}{\rho(\nu)} \int_0^\infty 
\frac{d\omega }{p^2+\omega^2}
\frac{1-e^{-\beta\omega}}{\tanh(\beta\omega/2)}
tr [\hat \rho(\nu) \hat J \hat \rho(\nu+\omega) \hat J]
\end{equation}
or, equivalently, 
\begin{equation}
\ell^2(p,\nu) = \frac{2}{\rho(\nu)} \int_{-\infty}^\infty d\omega 
\frac{1}{p^2+\omega^2}
tr [\hat \rho(\nu) \hat J \hat \rho(\nu+\omega) \hat J].
\end{equation}
These relations are obtained from the definition
Eq. (\ref{eq:L2pElowdens}) via Eq. (\ref{eq:Lextensive}).

The Bloch-Boltzmann 
theory is customarily obtained from the RTA by taking the 
noninteracting band eigenstates as the reference system.  
Within the present formalism this amounts to 
evaluating the trace in the
above equation assuming that the eigenstates of the Hamiltonian are good
momentum eigenstates, $|k\rangle$, with energy $E_k$. 
In that case $\langle k | \hat{J} |
k\rangle =v_k= dE_k/dk$ and we obtain 
\begin{equation}
  \label{eq:l2Bol}
  \ell^2(p,\nu) = \frac{2}{\rho(E_k)} \sum_k\frac{v_k^2}{p^2} \delta(\nu-E_k)
\end{equation}
Using Eqs. (\ref{eq:mobzero}) and  (\ref{eq:RTAbub}), with $\tau=1/p$ the relaxation time
for momentum eigenstates  
yields the Boltzmann form of the  mobility: 
\begin{equation}
  \label{eq:BolRTA}
  \mu=\frac{e}{k_B T} \langle  \tau v_k^2 \rangle
\end{equation}
where the thermal average is defined as  
$\langle  \tau v_k^2 \rangle= 
\frac{\sum_k e^{-\beta E_k} \tau v_k^2}{\sum_k e^{-\beta E_k}}$.


\begin{thebibliography}{99}
\bibitem{Podzorov} V. Podzorov, E. Menard, J. A. Rogers, M. E. 
Gershenson, Phys. Rev. Lett. 95, 226601 (2005).
\bibitem{Xie} H. Xie, H. Alves, A. F. Morpurgo, Phys. Rev. B 80,
245305 (2009).
\bibitem{Sakanoue} T.Sakanoue and H.Sirringhaus, Nat.Mater. 9, 736 (2010).
\bibitem{Liu} C. Liu, T. Minari, X. Lu, A. Kumatani, K. Takimiya, and
K. Tsukagoshi, Adv. Mater. 23, 523 (2011).
\bibitem{Minder} N. A. Minder, S. Ono, Z. Chen, A. Facchetti, and A.F.
Morpurgo, Adv. Mater. 24, 503 (2012).
\bibitem{Friedman} L. Friedman, Phys. Rev. 140, A1649 (1965).
\bibitem{Cheng} Y. C. Cheng et al., J. Chem. Phys. 118, 3764 (2003).
\bibitem{orgarpes11} S. Ciuchi and S. Fratini, Phys. Rev. Lett. 106,
  166403 (2011). 
\bibitem{Stojanovic} N. Vukmirovi\'c, C. Bruder and V. M. Stojanovi\'c, 
Phys. Rev. Lett. 109, 126407 (2012)
\bibitem{Munn} R. W.	Munn	and	R. J.	Silbey,	J. 
Chem.	Phys.	83, 1854 (1985).
\bibitem{Troisi06} A. Troisi \& G. Orlandi,  Phys. Rev. Lett. 96, 086601 (2006).
\bibitem{Picon} J.-D. Picon, M. N. Bussac, and L. Zuppiroli, Phys. Rev. B 75,
235106 (2007).
\bibitem{reconcile09} S. Fratini and S. Ciuchi, Phys. Rev. Lett. 103,
  266601 (2009). 
\bibitem{RTA11} S. Ciuchi, S. Fratini and D. Mayou,  Phys. Rev. B 83,
  081202(R) (2011). 
\bibitem{Karl} N. Karl, Organic Electronic Materials, 
edited by R. Farchioni and G. Grosso (Springer-Verlag, Berlin, 2001), 
pp. 283-326.
\bibitem{Fischer} M. Fischer, M. Dressel, B. Gompf, A.K. Tripathi,and
  J. Pflaum, Appl. Phys. Lett. 89, 182103 (2006). 
\bibitem{Li} Z. Q. Li, V. Podzorov, N. Sai, M. C. Martin, M. E. Gershenson,
M. DiVentra, and D. N. Basov, Phys. Rev. Lett. 99, 016403 (2007).
\bibitem{Hulea} I. N. Hulea, S. Fratini, 
H. Xie, C. L. Mulder, N. N. Iossad, G. Rastelli, S. Ciuchi, 
A. F. Morpurgo, Nat. Mater. , 5, 982 (2006).
\bibitem{Richards08} T. Richards, M. Bird, H. Sirringhaus, 
J. Chem. Phys.  128, 234905 (2008).
\bibitem{Kalb} W.L. Kalb, S. Haas, C. Krellner, T. Mathis, and B. Batlogg, 
Phys. Rev. B 81, 155315 (2010).
\bibitem{Chang11} J.-F. Chang, T. Sakanoue, Y. Olivier, T. Uemura,
M.-B. Dufourg-Madec, S. G. Yeates, J. Cornil, J. Takeya, A. Troisi and H. Sirringhaus, 
Phys. Rev. Lett. 107, 066601 (2011)
\bibitem{Mayou00} D. Mayou, Phys. Rev. Lett. 85, 1290 (2000).
\bibitem{Trambly06} G. Trambly de Laissardi\`ere, J.-P. Julien, and
  D. Mayou, Phys. Rev. Lett. 97, 026601 (2006).  
\bibitem{Trambly11} G. Trambly de Laissardi\`ere and
  D. Mayou, Mod. Phys. Lett. 25, 1019 (2011).  
\bibitem{Mahan} G. D. Mahan ``Many particle physics'',  second
  edition, Plenum Press N.Y. and London (1990).
\bibitem{symbols}
We are using the same symbols for the Fourier and Laplace
transforms for simplicity. These are however different 
functions and are implicitly
specified by their argument, $\omega$ or $p$ respectively.
\bibitem{Kubo57} R. Kubo, J. Phys. Soc. Japan. 12, 570 (1957). 
\bibitem{Coropceanu12} Y. Li, Y. Yi, V. Coropceanu and J.-L. Br\'edas,
  Phys., Rev. B 85, 245201 (2012).
\bibitem{Hatch} S. Ciuchi, R. C. Hatch, H. H\"ochst, C. Faber, X. Blase, and S. Fratini,
Phys. Rev. Lett. 108, 256401 (2012)  
\bibitem{LeeRMP} P. A. Lee and T. V. Ramakrishnan, Rev. Mod. Phys. 57,
  287 (1985).
\bibitem{vertex} The full neglect of vertex
corrections also  amounts to 
setting the transport scattering time  equal to the quasiparticle
lifetime \cite{Mahan}. 
In the band limit, this  leads to an overestimate of the
mobility by at most a  factor of two.
The multiplicative factor approaches one in the case where the transport
is dominated by incoherent tail states.
\bibitem{Fratini03} S. Fratini and S. Ciuchi, Phys. Rev. Lett. 91,
  256403 (2003). 
\bibitem{Bassler} H. B\"assler, Phys. Status Solidi B 175, 15 (1993).
\bibitem{Coehoorn} R. Coehoorn, W. F. Pasveer, P. A. Bobbert, and 
M. A. J. Michels, Phys. Rev. B 72, 155206 (2005).  
\bibitem{Thouless} D. J. Thouless, Phys. Rev. Lett. 39, 1167 (1977).
\bibitem{Cataudella} V. Cataudella, G. De Filippis, and C. A. Perroni,
  Phys. Rev. B 83, 165203 (2011).
\bibitem{TroisiAdv} A. Troisi,   Adv. Mat. 19, 2000 (2007).
\bibitem{Hannewald} K. Hannewald \& P. A. Bobbert. 
 Phys. Rev. B 69, 075212 (2004).
\bibitem{WangJCP07} L. J. Wang, Q. Peng, Q. K. Li, and Z. Shuai.
 J. Chem. Phys. 127, 044506 (2007).
 \bibitem{Wang} L. Wang, D. Beljonne, L. Chen and Q. Shi, 
 J. Chem. Phys. 134, 244116 (2011).
\bibitem{Ishii} H. Ishii, K Honma, N. Kobayashi and K. Hirose,
Phys. Rev. B  85, 245206 (2012).
\bibitem{HoHu} S. Fratini and S. Ciuchi, Phys. Rev. B 72, 235107 (2005).
\bibitem{Machida} S. Machida, et al., Phys. Rev. Lett. 104, 156401 (2010).
\bibitem{note:transient}The  transient localization
length is predicted to saturate to a constant value at very low temperatures,
because $L^2_{loc}(\tau_{in})$ is necessarily finite at any finite value of
$\tau_{in}$ as $T\to 0$ [see Eq. (\ref{eq:L2omega})].   This
implies a $\mu_e \propto T^{-1}$ dependence at low
temperature. This behavior however occurs 
beyond the limits of validity of 
our classical treatment for the molecular
vibrations, so that it is
not shown in Fig. \ref{fig:mobility}.
\bibitem{Marumoto} K. Marumoto, S. I. Kuroda, T.   Takenobu, and Y. Iwasa
  Phys. Rev. Lett. 97, 256603 (2006).
\bibitem{Matsui} H. Matsui, A. S. Mishchenko, and T. Hasegawa,
  Phys. Rev. Lett. 104, 056602 (2010).
  \bibitem{Mishchenko}
  A. S. Mishchenko, H. Matsui and T. Hasegawa,
Phys. Rev. B, 85, 085211 (2012).

\bibitem{Baym} G. Baym and L. P. Kadanoff, {\it "Quantum statistical mechanics: Green's function methods in equilibrium and nonequilibrium problems"} W. A: Benjamin, N.Y. (1962).
\bibitem{nota-ell-Kubo} The
  Thouless \cite{Thouless} definition adopted in
  Ref. \onlinecite{reconcile09} for the  
localization length differs in principle from the
definition Eq. (\ref{eq:Lextensive}) for two reasons. First, the
localization  length reported 
in Ref. \onlinecite{reconcile09}  was weighted by the
density of states  at energy $\nu$. Second, 
the present $\ell (\tau_{in},\nu)$ represents the {\it
  dynamical} localization of electronic states 
on a timescale $\tau_{in}$, while the
Thouless length is defined in the static limit $\tau_{in}\to \infty$.


\bibitem{Frisbie}
M. J. Panzer and C. D. Frisbie, Appl. Phys. Lett. 88, 203504 (2006)
\bibitem{Iwasa}
H. Shimotani, H. Asanuma, J. Takeya and Y. Iwasa, Appl. Phys. Lett 89, 
203501 (2006)









\end{thebibliography}
\end{document}